\documentclass{article}
\usepackage{ijcai21}

\usepackage{pdfpages}
\usepackage[noend]{algorithmic}
\usepackage{algorithm}
\usepackage{amsmath,epsfig,latexsym,graphics,subfigure}
\usepackage{booktabs} % For formal tables
\usepackage{graphicx}
\usepackage{url}
\usepackage{xcolor}
\usepackage{xspace}
\usepackage{caption}
\usepackage{enumitem}
\usepackage{subfigure}
\usepackage{amsthm}
\usepackage{dsfont}
\usepackage{colortbl}
\usepackage{multirow}
\usepackage{tabularx} % in the preamble
\usepackage{longtable}
\usepackage{epstopdf}
\usepackage{balance}
\usepackage{amssymb}

\usepackage{times}
\usepackage{soul}
\usepackage{url}
\usepackage[utf8]{inputenc}
\usepackage{amsmath}
\usepackage{amsthm}
\usepackage{booktabs}
\usepackage{algorithm}
\usepackage{algorithmic}
\usepackage{balance}  % for  \balance command ON LAST PAGE  (only there!
\definecolor{Gray}{gray}{0.85}
\newcommand{\cg}{\cellcolor{Gray}}
\newcounter{example}[section]
\renewcommand{\theexample}{\nthesection.\arabic{example}}
\newenvironment{example}{
     \refstepcounter{example}
     {\vspace{0.5ex} \noindent\bf  Example  \theexample:}}{
     \eop\vspace{0.5ex}} %\hspace*{\fill}\vspace*{1ex}}

\newcounter{definition}[section]
\renewcommand{\thedefinition}{\nthesection.\arabic{definition}}
\newenvironment{definition}{
     \refstepcounter{definition}
     {\vspace{0.5ex} \noindent\bf  Definition  \thedefinition:}}{
     \eop\vspace{0.5ex}} %\hspace*{\fill}\vspace*{1ex}}

\newcounter{property}[section]
\renewcommand{\theproperty}{\arabic{property}}
\newenvironment{property}{ \begin{em}
        \refstepcounter{property}
        {\vspace{0.5ex} \noindent \bf {Property \theproperty:}}}{
        \end{em}\vspace{0.5ex}} %\hspace*{\fill}\vspace*{1ex}}

\newcounter{theorem}[section]
\renewcommand{\thetheorem}{\nthesection.\arabic{theorem}}
\newenvironment{theorem}{\begin{em}
        \refstepcounter{theorem}
        {\vspace{0.5ex} \noindent\bf  Theorem  \thetheorem:}}{
        \end{em}\vspace{0.5ex}} %\hspace*{\fill}\vspace*{1ex}}

\newcounter{lemma}[section]
\renewcommand{\thelemma}{\nthesection.\arabic{lemma}}
\newenvironment{lemma}{\begin{em}
        \refstepcounter{lemma}
        {\vspace{0.5ex}\noindent \bf Lemma \thelemma:}}{
        \end{em}\vspace{0.5ex}} %\hspace*{\fill}\vspace*{1ex}}

\newcounter{remark}
\renewcommand{\theremark}{\Roman{remark}}
\newcounter{corollary}[section]
\renewcommand{\thecorollary}{\nthesection.\arabic{corollary}}
\newcommand{\myproof}{\noindent{\em Proof: }}

\newcommand{\nthesection}{\arabic{section}}
\newcommand{\eop}{\hspace*{\fill}\mbox{$\Box$}}
\newcommand{\stitle}[1]{\vspace{1ex} \noindent{\bf {#1}}}

\newcommand{\sstitle}[1]{\vspace{1ex} \noindent{\textit{ #1}}}

\newcommand{\kw}[1]{{\ensuremath {\mathsf{#1}}}\xspace}

\newcommand{\kwnospace}[1]{{\ensuremath {\mathsf{#1}}}}

\newcommand{\ei}{\end{itemize}}
\newcommand{\ee}{\end{enumerate}}

\newcommand{\beqn}{\begin{eqnarray*}}
\newcommand{\eeqn}{\end{eqnarray*}}

\newcounter{ccc}

\newcommand{\eat}[1]{}

\long\def\comment#1{}

\definecolor{lgray}{gray}{0.85}
\definecolor{llgray}{gray}{0.9}

\newcommand{\reffig}[1]{Figure~\ref{fig:#1}}
\newcommand{\refsec}[1]{Section~\ref{sec:#1}}
\newcommand{\reftable}[1]{Table~\ref{tab:#1}}
\newcommand{\refalg}[1]{Algorithm~\ref{alg:#1}}

\newcommand{\refdef}[1]{Definition~\ref{def:#1}}
\newcommand{\refthm}[1]{Theorem~\ref{thm:#1}}
\newcommand{\reflem}[1]{Lemma~\ref{lem:#1}}
\newcommand{\refex}[1]{Example~\ref{ex:#1}}

\newcommand{\refpro}[1]{Property~\ref{pro:#1}}

\newcommand{\topcaption}{%
 \setlength{\abovecaptionskip}{0pt}%
 \setlength{\belowcaptionskip}{-5pt}%
 \caption}

\newcommand{\baseline}{\kw{HOTDecom}}
\newcommand{\baselinep}{\kw{HOTDecom^+}}
\newcommand{\baselines}{\kw{HOTTopR}}

\newcommand{\hotddu}{\kwnospace{HOTD}\textrm{+}\kw{DU}}
\newcommand{\hotdd}{\kwnospace{HOTD}\textrm{+}\kw{D}}
\newcommand{\hotd}{\kw{HOTD}}
\newcommand{\hotdp}{\kw{HOTD^+}}
\title{Higher-Order Neighborhood Truss Decomposition}

\author{
Zi Chen$^1$
\and
Long Yuan$^2$
\and
Li Han$^1$
\and
Zhengping Qian$^3$
\affiliations
$^1$East China Normal University,
$^2$Nanjing University of Science and Technology,
$^3$Alibaba Group
\emails 
zchen,hanli@sei.ecnu.edu.cn,
longyuan@njust.edu.cn,
zhengping.qzp@alibaba-inc.com
}

\begin{document}

\maketitle

\begin{abstract}
$k$-truss model is a typical cohesive subgraph model and  has been received considerable attention recently. However, the $k$-truss model only considers the direct common neighbors of an edge, which restricts its ability to  reveal fine-grained structure information of the graph. Motivated by this, in this paper, we propose a new model named $(k, \tau)$-truss that considers the higher-order neighborhood ($\tau$ hop) information of an edge. Based on the $(k, \tau)$-truss model, we study the higher-order truss decomposition problem which computes the $(k, \tau)$-trusses for all possible $k$ values regarding a given $\tau$. Higher-order truss decomposition can be used in the applications such as community detection and search, hierarchical structure analysis, and graph visualization. To address this problem, we first propose a bottom-up decomposition paradigm in the increasing order of $k$ values to compute the corresponding $(k, \tau)$-truss. Based on the bottom-up decomposition paradigm, we further devise three optimization strategies to reduce the unnecessary computation.  We evaluate our proposed algorithms on real datasets and synthetic datasets, the experimental results demonstrate the efficiency, effectiveness and scalability of our proposed  algorithms.
\end{abstract}

\section{Introduction}
Graphs have been widely used to represent the relationships of entities in real-world applications \cite{DBLP:journals/pvldb/SahuMSLO17,DBLP:journals/pvldb/YuanQLCZ17,DBLP:journals/pvldb/OuyangYQCZL20}. With the proliferation of graph applications, plenty of research efforts have been devoted to cohesive subgraph models for graph structure analysis \cite{cohesivesubgraphbook}. Typical cohesive subgraph models include  clique \cite{luce1949method,DBLP:journals/tkde/YuanQZCY18,DBLP:journals/vldb/YuanQLCZ16,DBLP:conf/www/ChenY0QY20}, $k$-clique \cite{luce1950connectivity}, quasi-clique \cite{abello2002massive,pei2005mining}, $k$-core \cite{DBLP:conf/www/LiuYLQZZ19,seidman1983network,DBLP:journals/vldb/LiuYLQZZ20}, $k$-truss \cite{truss,DBLP:conf/dasfaa/WuYLYZ19}, and $k$-ECC \cite{DBLP:conf/edbt/ZhouLYLCL12,DBLP:conf/sigmod/ChangYQLLL13}.

Among them, the $k$-truss model is a typical cohesive subgraph model and has received considerable attention due to its unique cohesive properties on degree and bounded diameter \cite{DBLP:journals/pvldb/WangC12,DBLP:conf/sigmod/HuangCQTY14,DBLP:journals/pvldb/AkbasZ17,DBLP:conf/sigmod/LiuZ0XG20}. Given a graph $G$, for an edge $e = (u, v)$ in $G$, the support of $e$ is defined as the number of direct common neighbors of $u$ and $v$.  $k$-truss is the maximal subgraph $G'$ of $G$ such that the support of each edge in $G'$ is not less than $k -2$ \cite{truss}. Truss decomposition computes the $k$-truss in the graph for all possible $k$ values in $G$.  

\begin{figure}[t]
\begin{center}
\includegraphics[width=0.8\columnwidth]{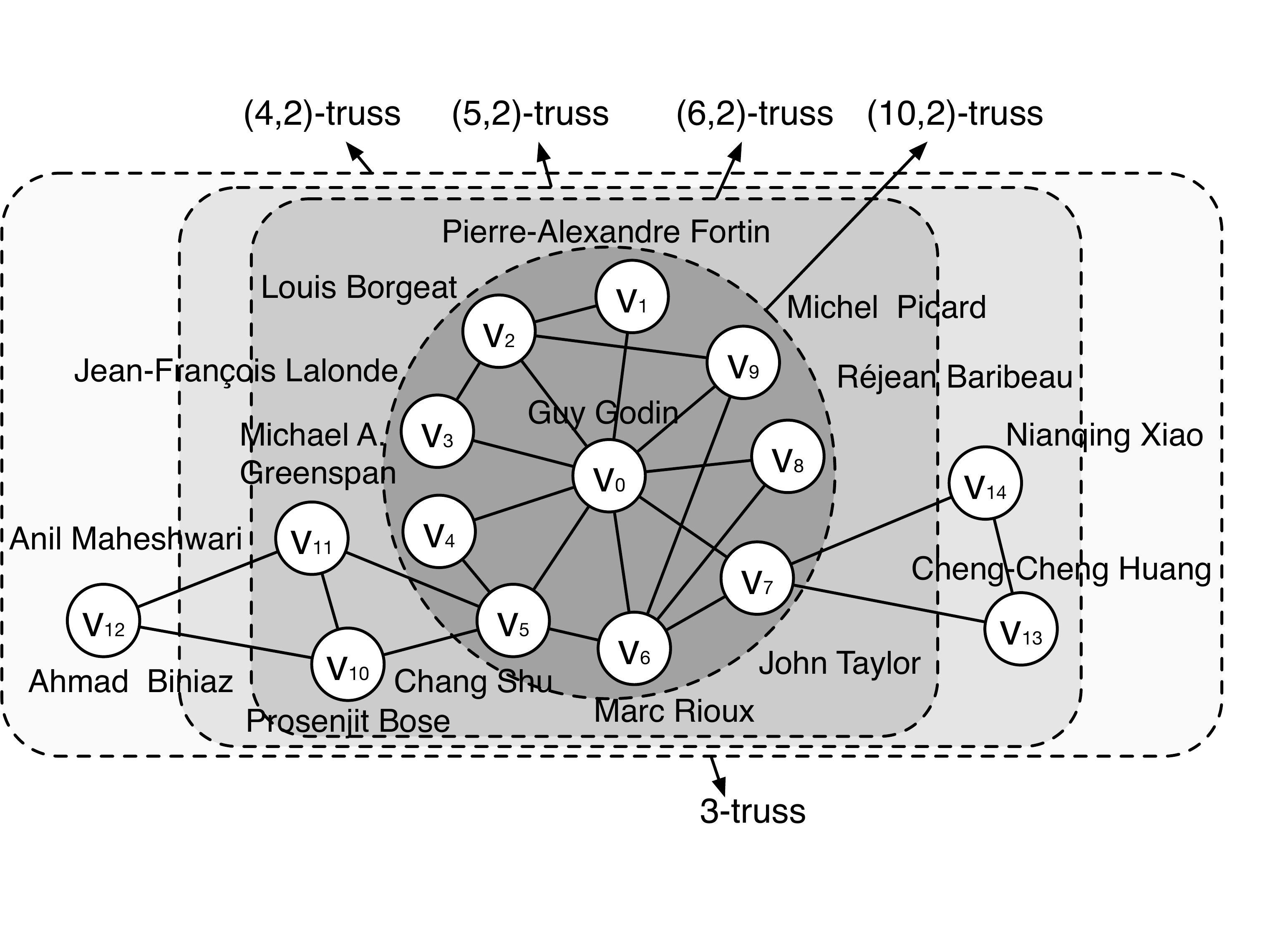}
\end{center}
\vspace{-0.9cm}
\caption{Part of the collaboration network in DBLP}
\label{fig:example_1}
%\vspace{-0.4cm}
\end{figure}

\stitle{Motivation.} Although the $k$-truss model and the corresponding truss decomposition have been successful in many applications,  the  model  lacks the ability to  reveal fine-grained structure information of the graph. Consider the graph  in \reffig{example_1}. \reffig{example_1} shows part of the collaboration network in the DBLP(\url{https://dblp.uni-trier.de/}), in which each node represents an author and each edge indicates the co-author relationship between two authors. The results of traditional truss decomposition are shown in \reffig{example_1}. The traditional truss decomposition treats the whole graph as a $3$-truss and is not able to provide more fine-grained structure information. 

On the other hand, the importance of the higher-order neighborhood (multiple-hop neighbors instead of direct neighbors) on the characterization of complex network  has been well established \cite{PhysRevE.73.046101,andrade2008characterization}, and remarkable results have been obtained in the network science due to the introduction of higher-order neighborhood \cite{DBLP:conf/icml/Abu-El-HaijaPKA19,DBLP:journals/corr/abs-1911-04129,DBLP:conf/sigmod/BonchiKS19,DBLP:conf/bigcomp/XueSS20,DBLP:conf/aaai/SunW0CDZQ20}.  Motivated by this, we propose the $(k, \tau)$-truss model by incorporating the higher-order neighborhood into the $k$-truss model. Formally, given a graph $G$ and an integer $\tau$, for an edge $e = (u, v)$, the higher-order support of $e$ is the number of $\tau$-hop common neighbors of $u$ and $v$. $(k, \tau)$-truss is the maximal subgraph $G'$ of $G$ such that the higher-order support of each edge in $G'$ is not less than $k-2$. Following the $(k, \tau)$-truss model, we study the higher-order truss decomposition  problem that computes the $(k, \tau)$-truss for all possible $k$ values in $G$ regarding a given $\tau$.

The benefits of the $(k, \tau)$-truss model are twofold: (1) it inherits the unique cohesive  properties on degree and bounded diameter of $k$-truss, which are shown in \refpro{degree} and \refpro{diameter} in \refsec{theo}. (2) It acquires the ability to reveal fine-grained structure information due to the introduction of higher-order neighborhood. Reconsider $G$ in \reffig{example_1}, \reffig{example_1} shows the higher-order truss decomposition results. By considering the higher-order neighborhood information, the whole graph(3-truss) can be further split into $(4,2)$-truss, $(5,2)$-truss, $(6,2)$-truss and $(10,2)$-truss. The hierarchy structure of the graph is clearly characterized by the  higher-order truss decomposition. Note that  a cohesive subgraph model named ($k$,$h$)-core that also considers higher-order neighborhood information is studied in \cite{khcore,DBLP:conf/sigmod/BonchiKS19}. However, the definition of our model is more rigorous, which makes our model can search ``core" of a ($k$,$h$)-core (e.g. a ($k$+1,$\tau$)-truss is a ($k$,$h$)-core but not vice versa,$\tau$=$h$). As shown in our case study (Exp-5), our model has higher ability to reveal fine-grained structure information than ($k$,$h$)-core. 

\stitle{Applications.} 
%As the higher-order truss inherits the cohesiveness of $k$-truss, 
Higher-order truss decomposition can be applied in the applications using the traditional $k$-truss decomposition since the $k$-truss model is a specific case of $(k, \tau)$-truss model when $\tau = 1$, namely $(k, 1)$-truss. These applications include community detection and search \cite{DBLP:conf/sigmod/HuangCQTY14,DBLP:journals/pvldb/AkbasZ17}. Moreover, as the higher-order truss decomposition can reveal more fine-grained structure of graphs compared with the traditional truss decomposition as shown in \reffig{example_1}, it can also be applied in the applications which focus on the hierarchical structure of graph, such like hierarchical structure analysis \cite{orsini2013evolution,DBLP:conf/sigmod/ShaoCC14,10.1371/journal.pone.0033799} and graph visualization \cite{colomer2013deciphering,JGAA-405,10.1007/3-540-45848-4_57}. Besides, due to consideration of the higher-order neighborhood with parameter $\tau$, users can control the cohesiveness of  decomposition result in a more flexible manner. 

\stitle{Challenges.} To conduct the higher-order truss decomposition, we first propose a bottom-up decomposition paradigm by extending the peeling algorithm for the traditional truss decomposition \cite{DBLP:journals/pvldb/WangC12}.  It conducts the higher-order truss decomposition in the increasing order of $k$ values. After computing the higher-order support for each edge, it iteratively removes  the edge $e$  with the minimum higher-order support in the  graph and updates the higher-order support of the edges whose higher-order support may be changed due to the removal of $e$ until the graph is empty.  

Although the peeling algorithm is suitable for the traditional truss decomposition, following the above bottom-up  decomposition paradigm directly cannot handle the higher-order truss decomposition efficiently. This is because, when an edge $e = (u, v)$ is removed, for the traditional truss decomposition,  we just need to decrease the support of edges $(u, w)$ and $(v, w)$  by 1, where $w$ is a common neighbor of $u$ and $v$. However, for the higher-order truss decomposition, when $e$ is removed, the scope of edges whose higher-order support may be changed due to the removal of $e$ is enlarged to all the edges incident to $u$, $v$ and their $\tau$-hop common neighbors. Moreover,  opposite to the traditional truss decomposition, we have no prior knowledge on the specific decreased value of the higher-order support of these edges. It  means the higher-order support of these edges has to be recomputed based on its definition instead of just decreasing by 1 as in traditional truss decomposition, which is prohibitively costly.  \emph{The enlarged scope of influenced edges} and \emph{the un-determination of the decreased higher-order support value} not only imply that the higher-order truss decomposition  is harder than the traditional truss decomposition, but also are the reasons why following the above bottom-up  decomposition paradigm directly  is inefficient for the higher-order truss decomposition.

\stitle{Our idea.} Revisiting the two reasons leading to the inefficiency of the  bottom-up  decomposition paradigm,  for the un-determination of the decreased higher-order support value,  it seems insoluable to obtain the higher-order support of an influenced edge without recomputation based on the definition. Hence, we focus on reducing the scope of influenced edges  whose higher-order support has to be recomputed for each removal of an edge.

To achieve this goal, we  follow the  bottom-up decomposition paradigm. We define the higher-order truss number of an edge as the maximal value of $k$ such that the edge is in the $(k, \tau)$-truss, but not in the $(k+1, \tau)$-truss.  When handling a specific $k$, we observe that for an edge with  higher-order truss number bigger than $k$, the correctness of its higher-order support in the remaining graph does not affect the correctness of the higher-order truss number computation for the edges whose higher-order truss number is $k$. It means that recomputing the higher-order support of the edges with higher-order truss number bigger than $k$ immediately after the removal of an edge is not necessary. Therefore, we propose a delayed update strategy and recompute the higher-order support when necessary. With this strategy, we can reduce the scope of the influenced edges whose higher-order support has to be recomputed. However, to fulfill this strategy, we have to know the higher-order truss number in prior, which is intractable. Consequently, we devise a tight lower bound of the higher-order truss number. When handling a specific $k$, we do not need to recompute the higher-order support for the edges whose lower bound of the higher-order truss number is bigger than $k$. 

Moreover, we further explore two optimization strategies, namely early pruning strategy and unchanged support detection strategy, to further reduce the scope of the influenced edges. Experiments on real datasets show that our improved algorithm can achieve up to 4 orders of magnitude  speedup compared with the baseline algorithm.

\stitle{Contributions.} We make the following contributions:

\begin{itemize}[leftmargin=*]
\item \sstitle{The first work to study the $(k, \tau)$-truss model.} Motivated by the traditional $k$-truss model ignores the higher-order neighborhood information of an edge, we propose the $(k, \tau)$-truss model. To the best of our knowledge, this is the first work considering the higher-order neighborhood information regarding the traditional $k$-truss model. Furthermore, we also prove the unique cohesive properties of the $(k, \tau)$-truss model.

\item \sstitle{Efficient algorithms for the higher-order truss decomposition.} We first devise a bottom-up decomposition paradigm by extending the peeling algorithm for traditional $k$-truss decomposition. Based on the bottom-up paradigm, we propose three optimization strategies to further improve the decomposition efficiency. Moreover, considering that some applications are more interested in the $(k, \tau)$-trusses with large $k$ values, we study the top $r$ $(k, \tau)$-trusses computation problem which returns the $(k, \tau)$-trusses with top $r$ $k$ values in the graph. We also propose an efficient algorithm for this problem.

\item \sstitle{Extensive performance studies on real datasets and synthetic datasets.} We conduct extensive experimental studies  on real datasets and synthetic datasets. For the efficiency, our improved algorithm can achieve up to 4 orders of magnitude speedup  compared with the direct bottom-up  decomposition paradigm. Besides, it also shows high effectiveness and scalability.  
\end{itemize}

\stitle{Outline.}  \refsec{pre} provides the problem definition. \refsec{theo} presents the theoretical cohesiveness properties of $(k, \tau)$-truss model. \refsec{baseline} introduces the bottom-up decomposition paradigm. \refsec{bu} presents our improved algorithms for higher-order truss decomposition problem. \refsec{td} presents our approach for top $r$ $(k, \tau)$-trusses computation problem which returns the $(k, \tau)$-trusses with top $r$ $k$ values in the graph. \refsec{exp} evaluates our algorithms and \refsec{related} reviews the related work. \refsec{con} concludes the paper. 

\section{Preliminaries}
\label{sec:pre}
Given an undirected and unweighted graph $G=(V,E)$, where $V(G)$ and $E(G)$ represent the set of vertices and the set of edges  in $G$, respectively, we use $n$ and $m$ to denote the number of vertices and the number of edges in $G$, i.e., $n=|V|$, $m=|E|$. In the graph $G$, a path is a sequence of vertices $p = (v_1,v_2,\cdots ,v_j)$ where $(v_i,v_{i+1}) \in  E$ for each $1 \leq i < j$ and cycles are allowed on $p$. Given a path $p$, the path length of $p$, denoted by $l(p)$, is the number of edges on $p$. The shortest path between two vertices $u$ and $v$ in $G$ is the path between these two vertices with the minimum length. We call the length of the shortest path between $u$ and $v$ is the distance of $u$ and $v$, and denote it as $\kw{dis}_G(u, v)$. Given a graph $G$, the diameter of $G$, denoted by $\omega(G)$,  is the maximum length of the shortest path between any pair of vertices in $G$. Given two vertices $u, v \in G$, $u$ is $\tau$-hop reachable from $v$, denoted by $v \rightarrow_{\tau} u$,  if there is a path $p$ between $u$ and $v$ with $l(p) \leq \tau$. Since we consider the undirected graph in this paper,  $v \rightarrow_{\tau} u$ if and only if $u \rightarrow_{\tau} v$. Given a vertex $v$, the $\tau$-hop neighbors of $v$, denoted by $N_\tau(v,G)$, is the set of vertices $u$ such that $u$ is $\tau$-hop reachable from $v$.  Similarly, the $\tau$-hop degree of $v$, denoted by $d_\tau(v,G)$, is the number of $\tau$-hop neighbors of $v$, i.e., $d_\tau(v,G) = |N_\tau(v,G)|$.

\begin{definition}
\label{def:neighbor}
\textbf{($\tau$-Hop Common Neighbor)} Given a graph $G$ and an integer $\tau$, for an edge $e = (u, v)$ in $G$, $w$ is a  $\tau$-hop  common neighbor of $e$ if $u \rightarrow_{\tau} w$ and $v \rightarrow_{\tau} w$.
\end{definition}

For an edge $e = (u, v)$, we use  $\Delta_\tau(e, G)$ to denote the set of $\tau$-hop common neighbors of $e$, i.e., $ \Delta_\tau(e, G) =   N_\tau(u,G) \cap N_\tau(v,G)$.

\begin{definition}
\label{def:sup}
\textbf{(Higher-Order Edge Support)} Given a graph $G$ and an integer $\tau$, for an edge $e$, the higher-order  support of $e$, denoted by $\kw{sup}_\tau(e,G)$, is the number of  $\tau$-hop common neighbors of $e$, i.e., $\kw{sup}_\tau(e,G)= |\Delta_\tau(e, G) |$.
\end{definition}

\begin{definition}
\label{def:truss}
\textbf{($(k,\tau)$-Truss).} Given a graph $G$ and an integer $\tau$, a $(k,\tau)$-truss is a maximal subgraph $G'$ of $G$ such that $\kw{sup}_\tau(e,G') \ge k-2$ for all $e \in E(G')$ and no more edges can be added into $G'$.
\end{definition}

\begin{definition}
\textbf{(Higher-Order Truss Number)}
\label{def:trussnum}
Given a graph $G$ and an integer $\tau$,  for an edge $e$, the higher-order truss number of $e$, denoted by $\phi_\tau(e, G)$, is the maximum value of $k$ such that $e$ is contained in the corresponding $(k, \tau)$-truss.
\end{definition}

\begin{figure}[t]
\begin{center}
\includegraphics[width=0.85\columnwidth]{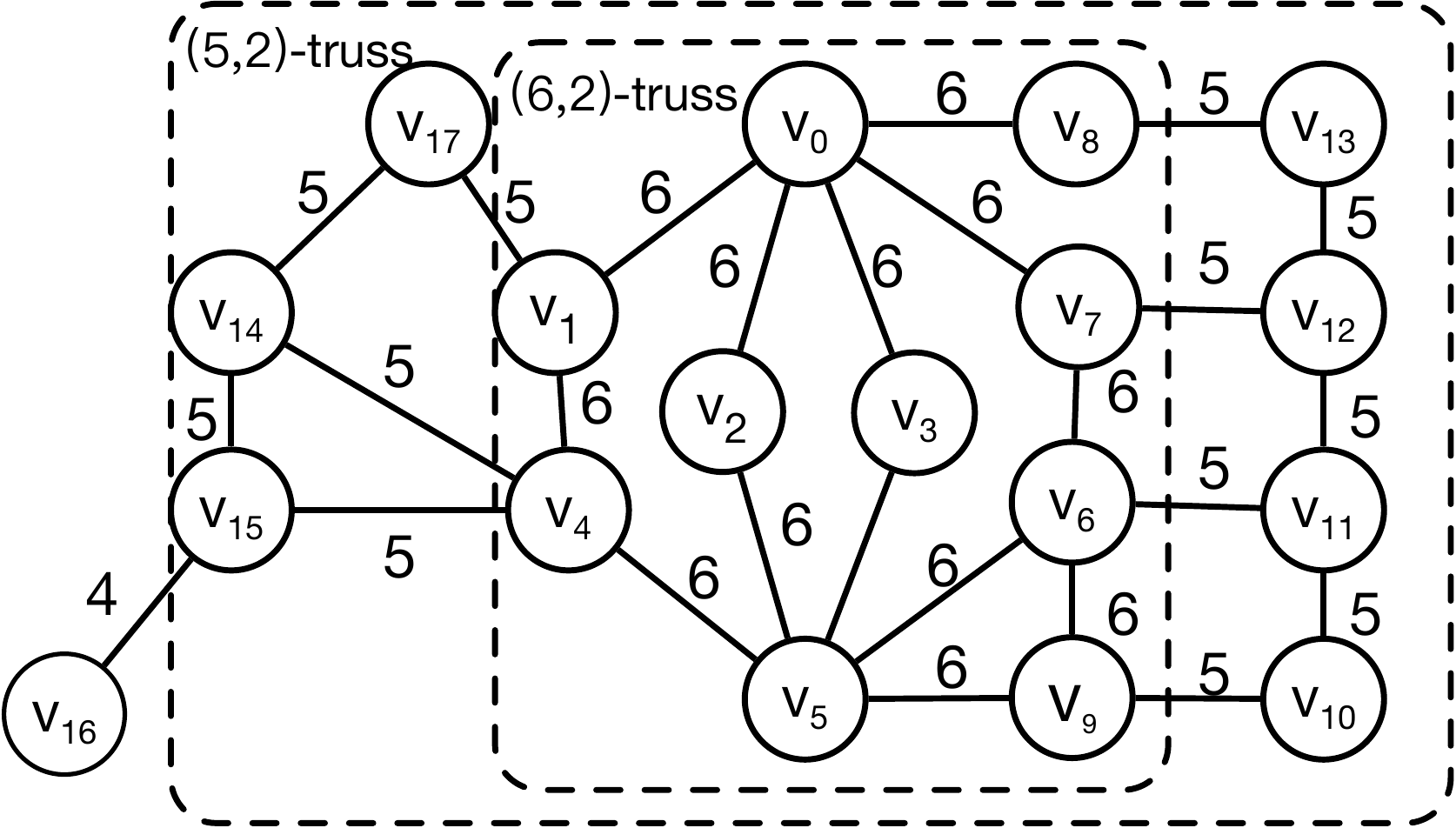}
\end{center}
%\vspace{-0.2cm}
\caption{An example of higher-order truss ($\tau=2$)}
\label{fig:example_2}
%\vspace{-0.4cm}
\end{figure}

\stitle{Problem Statement.} In the applications, the higher-order truss number of each edge is required. Given a graph $G$ and an integer $\tau$, in this paper, we study the higher-order truss decomposition problem which aims to compute  the  $(k,\tau)$-trusses of $G$ for all possible $k$ values regarding $\tau$. Straightforwardly, the $(k, \tau)$-truss of $G$ consists of the set of edges with higher-order truss number at least $k$, i.e., $\cup_{\phi_\tau(e, G) \geq k} e$. Therefore,  the higher-order truss decomposition  is equivalent to compute the higher-order truss number for each edge in $G$.

\begin{example}
Consider $G$  in \reffig{example_2} and assume $\tau=2$ (in the following examples, we always assume  $\tau = 2$). For $v_{0}$, its 2-hop neighbors $N_2(v_{0}, G) = \{v_1, \dots, v_8,v_{12},v_{13},v_{17}\}$. For $v_{1}$, its 2-hop neighbors $N_2(v_{1}, G) = \{v_0, v_2, \dots,v_5, v_7,v_8,v_{14},v_{15},v_{17}\}$. Therefore, $\Delta_2((v_{0}, v_{1}), G)=N_2(v_{0}, G)\cap N_2(v_{1}, G) = \{v_2, \dots v_5,v_7,v_8,v_{17}\}$, $\kw{sup}_2((v_0, v_1), G) = 7$.  $\kw{sup}_2(e, G)$ for each edge $e$ can be computed  similarly and we can find that $\kw{sup}_2(e, G) \geq 2$. Thus, $G$ itself is a $(4, 2)$-truss. Moreover, we can find that the subgraph $G'$ induced by $\{v_0, \dots, v_{15}\}$ is a $(5, 2)$-truss as $\kw{sup}_2(e, G') \geq 3$ for each edge $e$ in $G'$ and no more edges can be added to $G'$ to make it as a bigger $(5, 2)$-truss. $\phi_2((v_{15}, v_{16}), G)$ is 4 as $(v_{15}, v_{16})$ is contained in the $(4, 2)$-truss but not in the $(5, 2)$-truss.  The value of $\phi_2(e, G)$ is shown near each edge in \reffig{example_2}. The hierarchy structure of $G$ is clearly illustrated by the value of $\phi_2(e, G)$.

\end{example}

\section{Theoretical Properties of $(k, \tau)$-truss}
\label{sec:theo}

Although $\tau$-hop neighborhood is considered,  $(k, \tau)$-truss still have the cohesiveness properties on degree and diameter:

\begin{property}
\label{pro:degree}
\textbf{(Minimum Degree)} Given a $(k, \tau)$-truss $G'$ of $G$,  for each vertex $v \in V(G')$, $d_\tau(v, G') \geq k -1$.
\end{property}

\begin{proof} 
For a vertex $v$, let $e = (u, v)$ be an edge incident to $v$ in $G'$.  Based on \refdef{truss}, we have  $\kw{sup}_\tau(e, G') \geq k - 2$. According to \refdef{sup}, $v$ has at least $k - 2$ $\tau$-hop neighbors in $\Delta_\tau(e, G)$. Moreover, $u$ is a 1-hop neighbor of $v$. Thus,  $d_\tau(v, G') \geq k -1$.
\end{proof}
%\eop

\begin{property}
\label{pro:diameter}
\textbf{(Bounded Diameter)} Given a $(k, \tau)$-truss $G'$ of $G$, for each connected component $G''$ of $G'$, the diameter of $G''$ $\omega(G'') \leq \frac{2 \tau (|V(G'')|-1)}{k}$.
\end{property}
%\myproof 
\begin{proof} 
Without loss of generality,  let $p=(v_1, v_2,..., v_d)$ be the shortest path in $G''$ with $l(p) = \omega(G'')$. We use $V(p)$ to denote the set of vertices on $p$.  For an edge $e_i=(v_i,v_{i+1})$ on $p$, we use $\Gamma_i$ to denote the set of vertices in $\Delta_\tau(e_i, G'') \setminus V(p)$. According to \refdef{neighbor}, it is clear that  $|\Delta_\tau(e_i, G'') \cap V(p)| \leq 2\tau -2$.  Since $ \Gamma_i = \Delta_\tau(e_i, G'') \setminus (\Delta_\tau(e_i, G'') \cap V(p))$,  $|\Gamma_i| = |\Delta_\tau(e_i, G'')| - |\Delta_\tau(e_i, G'') \cap V(p)|$. According to \refdef{truss}, $|\Delta_\tau(e_i, G'')| \geq k -2$.  Together with $|\Delta_\tau(e_i, G'') \cap V(p)| \leq 2\tau -2$, we have $|\Gamma_i|\ge k-2\tau$.

Meanwhile, for a vertex $w \in V(G'')\setminus V(p)$, we have $|\{\Gamma_i:w \in \Gamma_i, 1 \leq i \leq d -1 \}| \leq 2 \tau$. This can be proved by contraction.  Assume that $|\{\Gamma_i:w \in \Gamma_i, 1 \leq i \leq d -1 \}| \geq 2\tau + 1$, let $E_\Gamma$ be the set of edges on $p$ such that $w$ is a $\tau$-hop common neighbor. Based on the assumption, $|E_\Gamma| \geq 2 \tau +1$. Let $e_j = (v_j, v_{j+1})$ and $e_k = (v_k, v_{k+1})$ be  two edges in $E_\Gamma$ such that the edges on $p$ between $e_j$ and  $e_k$ are maximum. We can derive that the length of the sub-path from $v_j$ to $v_{k+1}$ on $p$ is at least $2 \tau +1$ due to $|E_\Gamma| \geq 2 \tau +1$.  On the other hand,  since $w$  is a $\tau$-hop common neighbors of $e_j$ and $e_k$, there exists a path $p'$ from $v_j$ to $v_{k+1}$ through $w$ with length less than $2 \tau$. It contradicts with the assumption that $p$ is a shortest path from $v_1$ to $v_d$ as we can replace the sub-path from $v_j$ to $v_{k+1}$ on $p$ with $p'$ to obtain a shorter path. Therefore, $|\{\Gamma_i:w \in \Gamma_i, 1 \leq i \leq d -1 \}| \leq 2 \tau$.   Based on this, we can derive that $|\Gamma_1\cup \Gamma_2...\cup \Gamma_{d-1}| \ge \frac{(d-1)(k-2\tau)}{2\tau}$.  Since $V(G'') = \Gamma_1\cup \Gamma_2...\cup \Gamma_{d-1} \cup V(p)$, $|V(G'')| = |\Gamma_1\cup \Gamma_2...\cup \Gamma_{d-1}| + |V(p)|$, we can derive that $|V(G'')| \ge \frac{(d-1)(k-2\tau)}{2\tau} + d$. As $\omega(G'') = d -1$, $\omega(G'') \leq \frac{2 \tau (|V(G'')|-1)}{k}$, the property holds. 
%\eop
\end{proof}

\section{A Bottom-Up Decomposition Paradigm}
\label{sec:baseline}
In this section, we present a  bottom-up decomposition algorithm for higher-order neighborhood truss decomposition, which is  based on the following lemma:

\begin{lemma}
\label{lem:base}
Given a graph $G$ and an integer $\tau$, a $(k+1,\tau)$-truss is contained by a $(k,\tau)$-truss.
\end{lemma}

%\myproof 
\begin{proof}
	
According to \refdef{truss}, for each edge $e$ in a $(k+1, \tau)$-truss $G'$, $\kw{sup}_{\tau}(e, G') \geq k-1 \geq k-2$. It is clear that $G'$ is also a $(k, \tau)$-truss.
%\eop
\end{proof}

Based on \reflem{base}, for a given graph $G$, we can decompose $G$ in the increasing order of $k$. For a specific $k$, we compute the edges whose higher-order truss number is identical to $k$. As $(k+1, \tau)$-truss is contained in the $(k, \tau)$-truss, according to \refdef{truss}, we can remove these edges from $G$ and get the $(k+1, \tau)$-truss. We continue the procedure and  the higher-order truss number for each edge can be obtained when all the edges are removed. The pseudocode of the decomposition algorithm  is shown in \refalg{bu}

\stitle{Algorithm.} Following the above idea, the paradigm to conduct the higher-order truss decomposition, \baseline, is shown in \refalg{bu}.  Given a graph $G$ and an integer $\tau$, it first computes the higher-order  support for each edge in $G$ (line 1-2). Then, it  determines the higher-order truss number for each edge by iteratively removing the edges  until  $G$ is empty (line 3-9). Specifically, it first assigns the value of the minimum higher-order  support plus 2 among the edges in the remaining graph $G$ to $k$ (line 4). It means the remaining graph is at least a $(k, \tau)$-truss. Therefore, the higher-order truss number for the edges $e$ in the remaining graph with $\kw{sup}_{\tau}(e, G) \leq k -2$ is $k$ (line 6). After the higher-order truss number of $e$ is obtained, \refalg{bu} removes $e$ from $G$ (line 7). Due to the removal of $e$, the $\tau$-hop common neighbors of edges incident to vertices in $\{u\} \cup \{v\} \cup \Delta_{\tau}(e, G')$ could be changed in $G$. Consequently, \refalg{bu} recomputes the higher-order  support for these edges  with $\kw{sup}_\tau(e',G) > k-2$ (line 8-9). \refalg{bu} continues until $G$ is empty (line 3).

\begin{algorithm}[t]
\caption{ $\kw{HOTDecom}(G, \tau)$}
{
%\small
\begin{algorithmic}[1]
\label{alg:bu}
\FOR{\textbf{each} $e \in E(G)$}
\STATE compute $\kw{sup}_\tau(e,G)$;
%\STATE $i=\Delta_h(e,G)+2$; $B[i]\leftarrow B[i]\cup\{e\}$;
\ENDFOR

\WHILE{$G \neq \emptyset$}
\STATE $k \leftarrow \kw{min}_{e \in E(G)} \kw{sup}_\tau(e, G)$ + 2;
\WHILE{$\exists ~ e = (u, v) \in E(G)$ with $\kw{sup}_\tau(e, G) \leq k-2$}
\STATE $\phi_\tau(e, G) \leftarrow k$;
\STATE $G' \leftarrow G$; $G\leftarrow G\setminus e$;
\FOR{\textbf{each} $e' = (u',v')\in E(G)$ with $u', v'\in \{u\}\cup\{v\} \cup  \Delta_\tau(e, G') $ with $\kw{sup}_\tau(e',G) > k-2$}
\STATE compute $\kw{sup}_\tau(e',G)$;
\ENDFOR
\ENDWHILE
\ENDWHILE

\end{algorithmic}
}
\end{algorithm}

\begin{figure}
\begin{center}
\includegraphics[width=0.85\columnwidth]{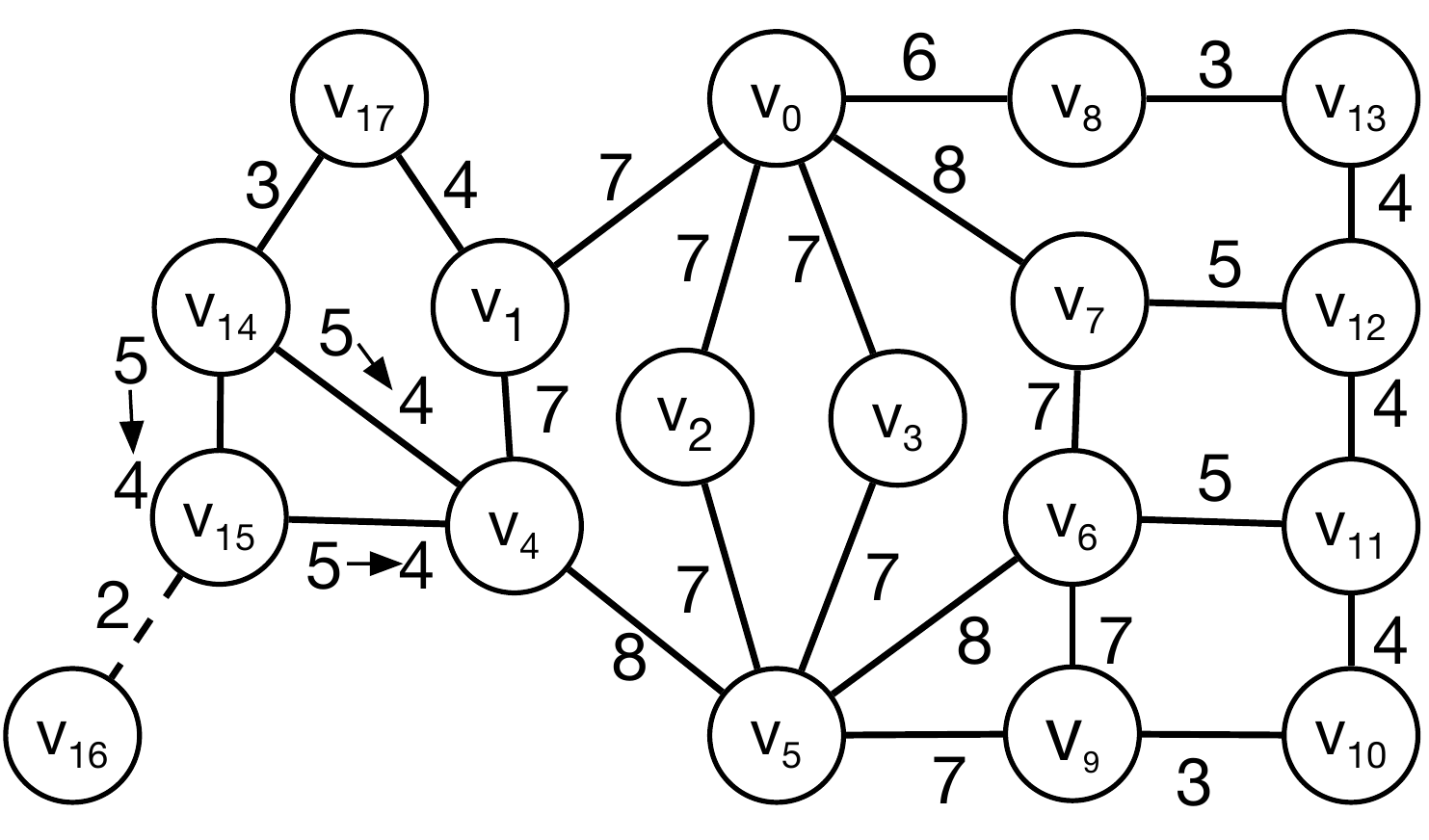}
\end{center}
%\vspace{-0.4cm}
\caption{The procedure of \baseline}
\label{fig:example_4}
%\vspace{-0.4cm}
\end{figure}

\begin{example}
\label{ex:baseline}
Reconsider the graph $G$ shown in \reffig{example_2}. \reffig{example_4} shows the procedure of \baseline to conduct the decomposition. It first computes  $\kw{sup}_2(e, G)$ for each edge, which is shown near each edge.  Since the minimum value of $\kw{sup}_2(e, G)$ among all the edges in $G$ is 2, then $k$ is assigned as 4 and $\phi_2((v_{15}, v_{16}), G)$ is 4. After that, $(v_{15}, v_{16})$ is removed and \baseline needs to update the higher-order support of $(v_{14}, v_{15})$. $\kw{sup}_2((v_{14}, v_{15}), G)$ decreases from 5 to 4, following $(v_4, v_{14})$ and $(v_4, v_{15})$. \baseline continues the above procedure until all the edges are removed. When it finishes,  the higher-order truss number for each edge is obtained.
\end{example}

\begin{theorem}
\label{thm:buc}
Given a graph $G$ and an integer $\tau$, \refalg{bu} computes the higher-order truss number for each edge  correctly.
\end{theorem}

%\myproof
\begin{proof}
	
 To show the correctness of \refalg{bu}, we only need to prove that the value assigned to $\phi_\tau(e, G)$ in line 6 of  \refalg{bu} is the correct higher-order truss number for every  edge $e \in E(G)$.

Based on the procedure of \refalg{bu}, for the first edge $e_1$ of $G$ processed in line 6, the value of $k$ is the correct higher-order truss number of $e_1$. This is because when  $e_1$ is processed,  $\kw{sup}_\tau(e, G)$ is correctly computed for every edge of $G$ in line 1-2 and the value of $k$ is the minimum value among all the computed $\kw{sup}_\tau(e, G)$ plus 2. Based on \refdef{truss},  the graph $G$ itself is a $(k, \tau)$-truss. Therefore, the higher-order truss number of $e_1$ is correctly computed.

Next, we show that the higher-order truss number is also correctly computed for the following processed edges $e_2, \dots, e_m$. We prove it by contradiction. Without loss of generality, let $e_i$ be the first edge assigned with wrong higher-order truss number, and the higher-order truss number assigned to $e_i$ by \refalg{bu} is $k_1$ while the correct higher-order truss number of $e_i$ is $k_2$. We first consider the case that $k_1 < k_2$. As $e_i$ is the first edge assigned with wrong higher-order truss number and the correct higher-order truss number of $e_i$ is $k_2 > k_1$, then, all the edges whose higher-order truss number is not less than $k_2$ have not been processed when processing $e_i$. According to \refdef{truss}, these edges together with $e_i$ consist of a $(k_2, \tau)$-truss and $\kw{sup}_\tau(e_i, G)$ is at least not less than $k_2 -2$.  Meanwhile \refalg{bu} assigns $k_1$ to $\phi_\tau(e_i, G)$ in line 6, it means $\kw{sup}_\tau(e_i, G)$ is less than $k_1 -2$ in line 5 before $e_i$ is processed in line 6.  It leads to the contradiction against the assumption that $k_1 < k_2$. Therefore, $k_1 < k_2$ is impossible.   Similarly, we can prove that $k_1 > k_2$ is also impossible. Therefore, the higher-order truss number is also correctly computed for $e_2, \dots, e_m$. Combining these two cases together, the theorem holds.
%\eop
\end{proof}

\begin{theorem}
\label{thm:butime}
Given a graph $G$ and an integer $\tau$, the time complexity of \refalg{bu} is $O(m\cdot m_\tau \cdot (m_\tau+n_\tau))$, where $m_\tau$ and $n_\tau$ are the maximum numbers of edges and vertices within the $\tau$-hop neighborhood in the graph, respectively.
\end{theorem}

%\myproof
\begin{proof}
	
 In \refalg{bu}, we first compute $\kw{sup}_\tau(e, G)$ for each edge in $G$ in line 1-2. To compute $\kw{sup}_\tau(e, G)$ for an edge $e = (u, v)$, we first retrieve $N_\tau(u, G)$ and $N_\tau(v, G)$ of $u$ and $v$ by breadth-first search, which costs $O(m_\tau+n_\tau)$ time. Then, the computation of $\Delta_\tau(e, G)$ costs $O(n_\tau)$ time. Therefore, the computation of line 1-2 can be finished in $O(m \cdot (m_\tau+n_\tau))$. For line 3-9, when an edge $e$ is removed in line 7, $O(m_\tau)$ edges in its $\tau$-hop neighborhood need to update their higher-order edge support and each update consumes $O(m_\tau+n_\tau)$ time. Therefore, the time for updating higher-order edge support in line 8-9 can be bounded by $O(m_\tau \cdot (m_\tau+n_\tau))$. In \refalg{bu}, each edge is removed once in line 7 and line 4-5 can be finished in const time by using a bin array.  As a result, the time complexity of line 3-9 can be bounded by $O(m\cdot m_\tau \cdot (m_\tau+n_\tau))$. Therefore, the overall time complexity of \refalg{bu} is $O(m\cdot m_\tau \cdot (m_\tau+n_\tau))$. %Obviously, the most calculation load is to compute and update the higher-order edge support of edges in $G$.
%\eop
\end{proof}

\section{An Improved Decomposition Algorithm}
\label{sec:bu}
%To address the ($k$,$h$)-truss decomposition proplem, in this section, we first propose our approach in the bottom-up fashion, intuitively, we want to decompose the $(k,h)$-truss by increasing $k$. Then, to further improve the efficiency of our algorithm, we design a significant lower bound to reduce the calculation load.
% We first propose a baseline algorithm and show its computation limited. Then,
In this section, we aim to  improve the performance of the bottom-up decomposition paradigm. According to \refthm{butime}, the most time-consuming part of \refalg{bu} is the higher-order support update in line 8-9. Compared with the time complexity of line 1-2, an additional term $m_\tau$ is introduced in that of line 3-9.  When an edge $e$ is removed from $G$ in line 7, \baseline updates $\kw{sup}_{\tau}(e', G)$ for all the edge $e'$ incident to vertices in  $\{u\} \cup \{v\} \cup \Delta_{\tau}(e, G')$  in line 8-9 immediately. The immediate update strategy adopted by \baseline leads to the term $m_\tau$ in \refthm{butime}, which makes it prohibitively costly.

%In this section, we explore opportunities to reduce the unnecessary higher-order support update  to improve the efficiency.

To address the performance issue in \baseline, we explore three optimization strategies, namely delayed update strategy, early pruning strategy, and unchanged support detection strategy  to avoid unnecessary higher-order support update. In this section, we first show these three proposed strategies. Then, we present our improved algorithm for the higher-order truss decomposition.

%\subsection{Three Optimization Strategies}

\subsection{A Delayed Update Strategy}
%\stitle{\underline{A Delayed Update Strategy.}}
\label{sec:delay}
Since the immediate update strategy leads to the term $m_\tau$ in the time complexity of \baseline,  our first idea is to explore the opportunities to delay the higher-order support update until necessary. To achieve this goal, we first define:

\begin{definition}
\textbf{(Truss Number Bounded Edge Set)} Given a graph $G$, an integer $\tau$,  and a condition $f(\phi)$, the truss number bounded edge set, denoted by $\Phi_{\tau, f(\phi)}(G)$, is the set of edges whose higher-order truss number $\phi_\tau(e, G)$ satisfies $f(\phi)$.
\end{definition}

%to further improve the efficiency of \bus algorithm, we need to reduce the amount of $\Delta_h(e,G)$ updation.

Revisiting \refalg{bu}, the key point to guarantee the correctness of the \baseline is that the value of $\kw{sup}_\tau(e, G)$ for an edge $e \in \Phi_{\tau, \phi = k}(G)$ must be not greater than $k-2$ when handling a specific $k$ in line 5-6. Following this, \baseline updates $\kw{sup}_{\tau}(e', G)$ for the edges incident to vertices in $\{u\} \cup \{v\} \cup \Delta_{\tau}(e, G')$ to keep the invariant.  On the other hand, according to \refdef{truss}, when handling a specific $k$, for an edge $e'' \in \Phi_{\tau, \phi > k}(G)$, it is not necessary to update the value of $\kw{sup}_\tau(e'', G)$  immediately based on the following two reasons: (1) for the edge $e \in \Phi_{\tau, \phi = k}(G)$, the value of $\kw{sup}_\tau(e'', G)$ does not affect the correctness of computing $\phi_{\tau}(e, G)$ in \refalg{bu} as $\kw{sup}_\tau(e'', G) \geq \phi_{\tau}(e'', G)-2 > k-2$.   (2) For the edge $e''$ itself, we can delay the computation of $\kw{sup}_{\tau}(e'', G)$ until handling $k$ whose value is identical to $\phi_\tau(e'', G)$  as $\phi_\tau(e'', G)$ is determined only by $\Phi_{\tau, \phi \geq k}$.

Following this idea, assume that we have already known the higher-order truss number of edges in prior, then, when an edge $e$ is removed in line 7 of \refalg{bu}, instead of updating $\kw{sup}_{\tau}(e', G)$ for the edge $e'$ incident to vertices in $\{u\} \cup \{v\} \cup \Delta_{\tau}(e, G')$, we only need to update $\kw{sup}_{\tau}(e', G)$ for edges with $\phi_\tau(e') = k$. With this \emph{delayed update strategy}, we can significantly reduce the number of edges updating their  values of $\kw{sup}_\tau(e', G)$ in line 9, which improves the performance of \baseline consequently.

\stitle{Lower bound of $\phi_\tau(e, G)$.} 
However, it is intractable to obtain the higher-order truss number directly before the decomposition as it is our goal. Despite the intractability, for an edge $e'$, if we can obtain a lower bound $\underline{\phi}_\tau(e', G)$ of its higher-order truss number rather than the exact value, when handling a specific $k$, we only update $\kw{sup}_{\tau}(e', G)$ for edges with $\underline{\phi}_\tau(e', G) \leq k$, the above inference still establishes. Therefore, the remaining problem is how to obtain a tight lower bound for an edge efficiently. To achieve this goal, we have the following lemma:

%we can still achieve the goal by only updating $\kw{sup}_{\tau}(e', G)$ for edges with $\underline{\phi}_\tau(e', G) \leq k$. 

\begin{lemma}
\label{lem:subgraph}
Given a graph $G$ and an integer $\tau$, let $G'$ be a subgraph of $G$, for any edge $e \in E(G')$, $\phi_{\tau}(e, G) \geq \phi_{\tau}(e, G')$.
\end{lemma}

\begin{proof}
This lemma can be proved directly based on \refdef{truss}.
\end{proof}

According to \reflem{subgraph}, for an edge in a graph $G$, the higher-order truss number of the edge in any subgraph of $G$ is a lower-bound of the higher-order truss number of the edge in $G$. Moreover, we have the following lemma:

\begin{lemma}
\label{lem:lb}
Given a graph $G$, for a subgraph $G'$ in $G$, if the diameter $\omega(G') \leq \tau$, then $G'$ is a $(|V(G')|, \tau)$-truss.
\end{lemma}

\begin{proof}	
Since the diameter $\omega(G') \leq \tau$, each vertex in $G'$ is $\tau$-hop reachable with each other. It means that for an edge $e = (u, v)$ in $G'$, the $\tau$-hop common neighbors of $e$ in $G'$ are all other vertices in $G'$ except $u$ and $v$. Therefore, we can drive that $\kw{sup}_{\tau}(e, G') = |V(G')| -2 $ for all the edges in $G'$.  According to \refdef{truss},  $G'$ is a $(|V(G')|, \tau)$-truss. The lemma holds.
\end{proof}

According to  \reflem{subgraph} and \reflem{lb}, for a  graph $G$ and an integer $\tau$, if we have a subgraph $G'$ of $G$ such that $\omega(G') \leq \tau$, then, for any edge $e \in E(G')$, we have $\phi_\tau(e, G) \geq |V(G')|$. Based on this, we define:

\begin{definition}
\label{def:tdiameter}
\textbf{(Vertex Centric $\tau$-Diameter Subgraph)} Given a graph $G$ and an integer $\tau$, for a vertex $v \in V(G)$, the vertex centric $\tau$-diameter subgraph of $v$, denoted by $\Omega(v)$, is the subgraph of $G$ induced by the set of vertices in $\{v\}\cup \{w | w \rightarrow_{\lfloor \frac{\tau}{2}\rfloor} v\}$.
\end{definition}

\begin{lemma}
\label{lem:tdiameter}
Given a vertex $v$ in a graph $G$  and an integer $\tau$, for an edge $e = (u, v) \in E(G)$, $\phi_\tau(e, G) \geq |V(\Omega(v))|$.
\end{lemma}

\begin{proof}
We consider two cases: (1) $\tau$ is even. Based on \refdef{tdiameter}, $\omega(\Omega(v)) \leq \tau$. According to \reflem{lb}, $\Omega(v)$ is a $(|V(\Omega(v)|, \tau)$-truss. Since $\Omega(v)$ is a subgraph of $G$, according to \reflem{subgraph},  $\phi_\tau(e, G) \geq |V(\Omega(v))|$. (2) $\tau$ is odd. Based on \refdef{tdiameter}, $\omega(\Omega(v)) \leq \tau -1 $. According to \reflem{lb}, $\Omega(v)$ is a $(|V(\Omega(v)|, \tau-1)$-truss. As $\Omega(v)$ is a subgraph of $G$, according to \reflem{subgraph},  $\phi_{\tau-1}(e, G) \geq |V(\Omega(v))|$. Following \refdef{truss}, $\phi_{\tau}(e, G) \geq \phi_{\tau-1}(e, G)$. Thus, $\phi_\tau(e, G) \geq |V(\Omega(v))|$. Combining the above two cases, the lemma holds.
\end{proof}

Following \reflem{tdiameter}, for an edge $e = (u, v) \in E(G)$,  we can derive that $\phi_\tau(e, G) \geq |V(\Omega(u))|$ as well. Moreover, we can also derive that:

\begin{lemma}
\label{lem:wdiameter}
Given a graph $G$ and an integer $\tau$, for an edge $e = (u, v) \in E(G)$,  $\phi_\tau(e, G) \geq  \kw{max}\{|V(\Omega(w))|\}$, where $w \in \Delta_{\lfloor \frac{\tau}{2} \rfloor}(e, G)$.
\end{lemma}

\begin{proof}
This lemma can be proved similarly as \reflem{tdiameter}.
\end{proof}

According to \reflem{tdiameter} and \reflem{wdiameter}, for an edge $e = (u, v)$, a lower bound of $e$ is $\kw{max}\{|V(\Omega(u))|$, $|V(\Omega(v))|, |V(\Omega(w))|\}$, where $w \in \Delta_{\lfloor \frac{\tau}{2} \rfloor}(e, G)$. However, in this case, if $\tau$ is odd, the value of $\underline{\phi}_{\tau}(e, G)$ is identical to that of $\underline{\phi}_{\tau-1}(e, G)$. To address this problem, we  define:

\begin{definition}
\label{def:edgetdiameter}
\textbf{(Edge Centric $\tau$-Diameter Subgraph)} Given a graph $G$ and an odd integer $\tau$, for an edge $e = (u, v) \in E(G)$, the edge centric $\tau$-diameter subgraph of $e$, denoted by $\Omega(e)$, is the subgraph of $G$ induced by the set of vertices in  $\{u\} \cup \{v\} \cup \{w| w \rightarrow_{\lfloor \frac{\tau}{2}\rfloor} u \vee  w \rightarrow_{\lfloor \frac{\tau}{2}\rfloor} v\}$.
\end{definition}

\begin{lemma}
\label{lem:edgediameter}
Given an edge $e$ in a graph $G$  and an odd integer $\tau$,  $\phi_\tau(e, G) \geq |V(\Omega(e))|$.
\end{lemma}

\begin{proof}
The lemma can be prove similarly as \reflem{tdiameter}.
\end{proof}

Therefore, our lower bound is defined as follows:

\begin{definition}
\label{def:lb}
\textbf{(Lower Bound $\underline\phi_\tau(e, G)$ )}  Given a graph $G$ and an integer $\tau$, for an edge $e = (u, v) \in E(G)$,  $\underline{\phi}_\tau(e, G) =$

$$
 \begin{cases}
~\kw{max}\{|V(\Omega(u))|, |V(\Omega(v))|, |V(\Omega(w))|\},& \tau~is~ even\\
~\kw{max}\{|V(\Omega(e))|, |V(\Omega(w))|\},& \tau ~is~ odd,\\
\end{cases}$$
 where $w \in \Delta_{\lfloor \frac{\tau}{2} \rfloor}(e, G)$.
\end{definition}

\begin{figure}
\begin{center}
\begin{tabular}[t]{c}
\subfigure[\small{$\tau$ is even}]{
\label{fig:subg}
\includegraphics[width=0.45\columnwidth]{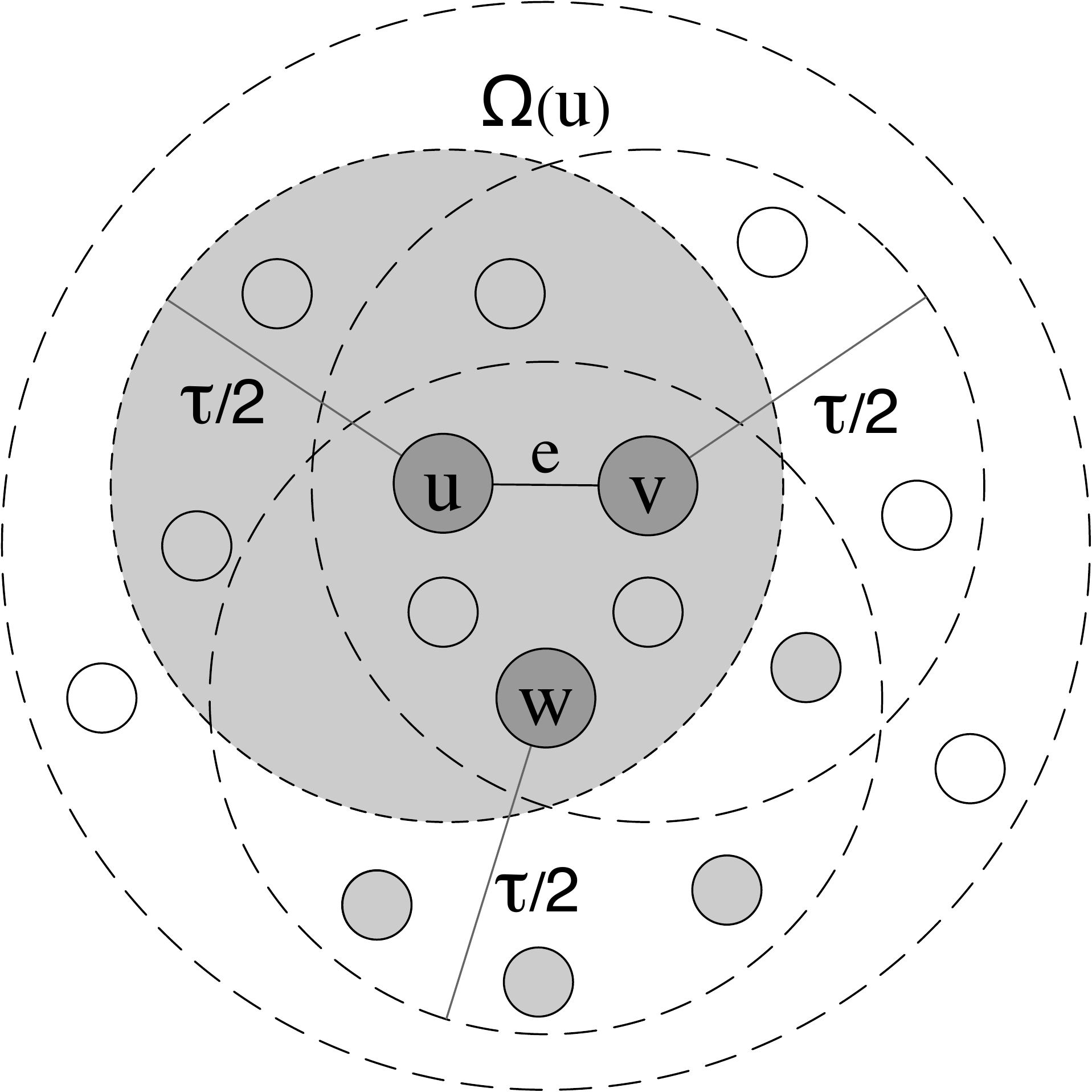}}
\subfigure[\small{$\tau$ is odd}]{
\label{fig:lbb}
\includegraphics[width=0.45\columnwidth]{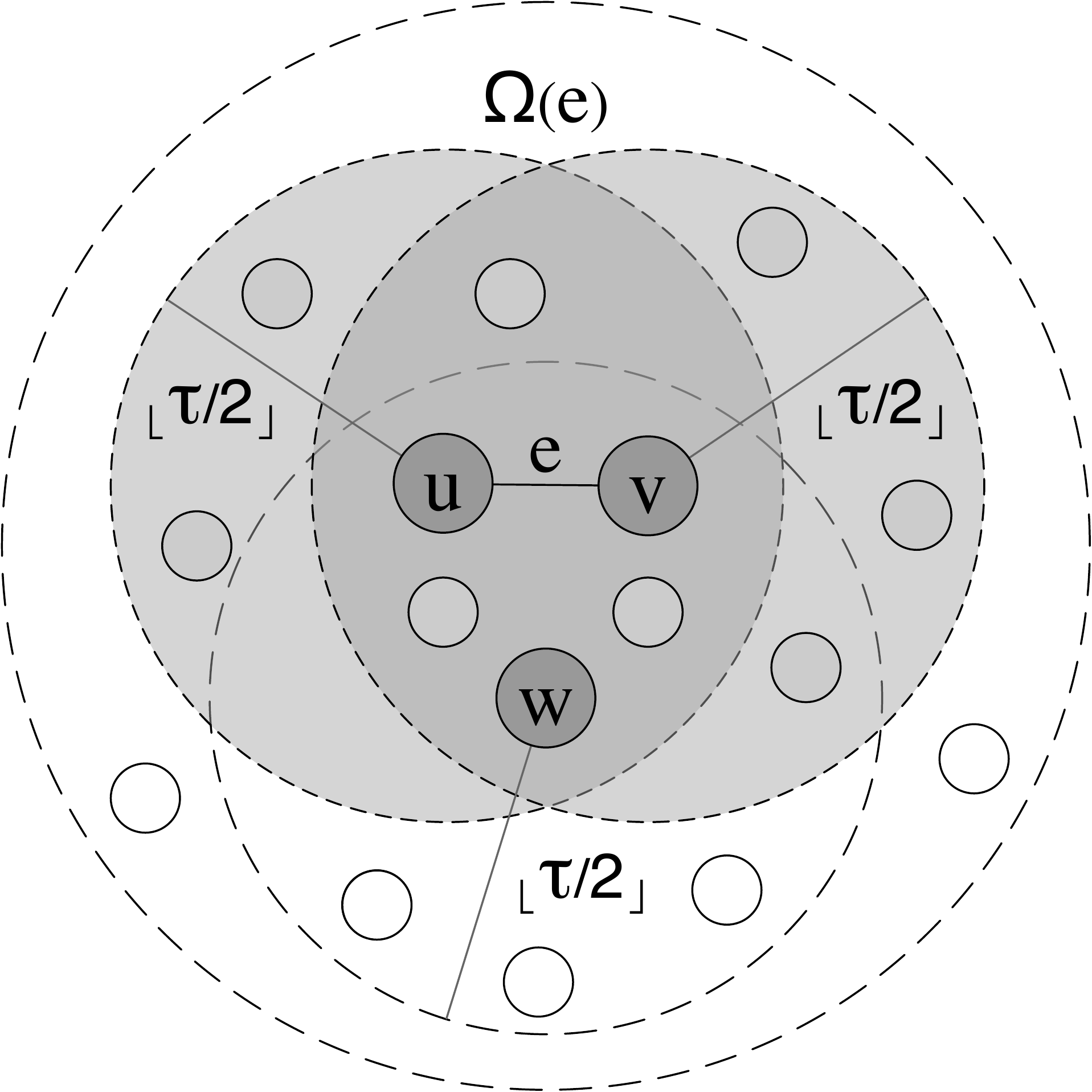}}
\end{tabular}\end{center}
\vspace{-0.4cm}
\caption{$\underline{\phi}_\tau(e, G)$ for an edge $e = (u, v)$}
\label{fig:lb}
%\vspace{-0.6cm}
\end{figure}

\begin{example}
\reffig{lb} illustrates the idea of $\underline{\phi}_\tau(e, G)$ for an edge $e = (u, v)$. In \reffig{lb} (a),  the shadowed dashed circle represents the vertex centric $\tau$-diameter subgraph $\Omega(u)$. In \reffig{lb} (b), the shadowed part represents the edge centric $\tau$-diameter subgraph $\Omega(e)$. When $\tau$ is even, we choose $\kw{max}\{|V(\Omega(u))|$,$|V(\Omega(v))|$,$|V(\Omega(w))|\}$ as $\underline{\phi}_\tau(e, G)$. When $\tau$ is odd, we choose $\kw{max}\{|V(\Omega(e))|, |V(\Omega(w))|\}$ as $\underline{\phi}_\tau(e, G)$. For a concrete example, consider the graph $G$ shown in \reffig{example_2}, take the edge $(v_0, v_1)$ as an example,  $\tau = 2$, since $\tau$ is even, we just need to consider $\Omega(v_0) = \{v_0, v_1,v_2,v_3,v_7,v_8\}$, $\Omega(v_1) = \{v_0,v_1,v_4,v_{17}\}$. Therefore, $\underline{\phi}_2((v_0, v_1), G) = \kw{max}\{|\Omega(v_0)|, |\Omega(v_1)|\} = 6$.
\end{example}

\subsection{An Early Pruning Strategy}
%\stitle{\underline{An Early Pruning Strategy.}}
With the lower bound, we can delay the computation of higher-order support when necessary, which consequently reduces the number of higher-order support update in line 8-9 of \refalg{bu}. On the other hand, when handling a specific $k$ during decomposition, if we have  some lightweight methods that can determine the higher-order truss number of an edge $e$ cannot be larger than $k$, then we can directly obtain  $\phi_\tau(e, G) = k$, which means we can correctly obtain the higher-order truss number of $e$ without the computation of its higher-order support. As a result, we can still reduce the number of higher-order support update and improve the performance of \baseline. Following this idea, we have:

\begin{lemma}
\label{lem:vertexprune}
Given a graph $G$, two integers $k$ and $\tau$, for an edge $e = (u, v) \in V(G)$, if $d_\tau(u, G)\le k-1 \vee d_\tau(v, G)\le k-1$, $\phi_\tau(e, G) \leq k$.
\label{lem:prun}
\end{lemma}

\begin{proof}
According to \refpro{degree}, if $d_\tau(u, G)\le k-1$ (resp. $d_\tau(v, G)\le k-1)$, $u$ (resp. $v$) is not contained in a $(k+1, \tau)$-truss. Thus, $e$ is not contained in a $(k+1, \tau)$-truss. Based on \refdef{trussnum}, $\phi_\tau(e, G) \leq k$.
\end{proof}

With \reflem{vertexprune}, when handling a specific $k$, we can prune an edge by computing the $\tau$-hop degree of its two incident vertices. Moreover, if we have computed $d_\tau(v, G)$ for a vertex $v$ and the value of $d_\tau(v, G)$ is not greater than $k -1$, then  all the edges incident to $v$ can be pruned. In other words, by computing the $\tau$-hop degree of a single vertex, we can reduce the computation of higher-order support for multiple edges, which achieves the goal of early pruning.

\subsection{An Unchanged Support  Detection Strategy}
In addition to the above discussed optimization strategies, here, we aim to explore the edges  whose higher-order  support unchanged after the removal of an edge to avoid the update. When an edge $e = (u, v)$ is removed , we have:

\begin{lemma}
\label{lem:unchanged}
Given an edge $e = (u, v)$ in a graph $G$, let $G'$ be the graph after the removal of $e$, for an edge $e' = (u', v')$  in $G'$ where $u', v' \in \Delta_\tau(e, G)$,  if the distance between $u$ (resp. $v$) to $u'$ (resp. $v'$) in $G$ and $G'$ keep the same, then, $\kw{sup}_\tau(e', G)  = \kw{sup}_\tau(e', G')$.
\end{lemma}

\begin{proof}
According to \refdef{sup}, we can prove the lemma if we can  prove that $N_\tau(u', G) = N_\tau(u', G')$ and $N_\tau(v', G) = N_\tau(v', G')$. Therefore, we first prove $N_\tau(u', G) = N_\tau(u', G')$. It can be proved by contradiction. Assume that  $N_\tau(u', G)$ and $N_\tau(u', G')$ are different due to the removal of $e'$. Since the removal of an edge only leads to the reduction of $\tau$-hop reachable vertices from a specific vertex, we can derive that  $N_\tau(u', G') \subseteq N_\tau(u', G)$. Without lose of generality, let $w$ be a vertex in $N_\tau(u', G) \setminus N_\tau(u', G')$. Then, we can derive $u' \rightarrow_{\tau}  w$ in $G$ but $u' \nrightarrow_{\tau} w$ in $G'$  according to \refdef{sup}. Therefore, the shortest path from $u'$ to $w$ must pass through $(u, v)$ in $G$.  Let the shortest path from $u'$ to $w$ be $p = (u', \dots, u, v, \dots, w)$. Since the shortest path from $u'$ to $w$ passes through $(u, v)$, we can derive that the shortest path from $u'$ to $v$ is the sub-path $p' = (u', \dots, u, v) $ of $p$ and $l(p')  < l(p'')$, where $p''$ is other arbitrary path from $u'$ to $v$ not passing through $(u, v)$ in $G$. Therefore, after the removal of $e$, the distance between $u'$ and $v$ must be larger than $l(p')$ in $G'$, which contradicts with the fact that distance between $u'$ and $v$ in $G$ and $G'$ are the same. Therefore, $N_\tau(u', G) = N_\tau(u', G')$. Similarly, we can prove $N_\tau(v', G) = N_\tau(v', G')$. Therefore,  $\kw{sup}_\tau(e', G)  = \kw{sup}_\tau(e', G')$.
\end{proof}

According to \reflem{unchanged}, for an edge $(u, v)$ in $G$, we can maintain the distance between $u$ (resp. $v$) and the vertices in $\Delta_\tau(e, G)$ before and after the removal of $(u, v)$. For those vertices $V' \subseteq \Delta_\tau(e, G)$ such that the distance between them and $u$ (resp. $v$) keep the same, we can guarantee that the higher-order  support of the edge $e'$ connecting any two vertices in $V'$ is unchanged after the removal of $(u, v)$.   Compared with updating the higher-order support for these edges $e'$ directly,  the above distance can be obtained   by just two BFS traversals starting from $u$ and $v$, respectively, which means we can further reduce the number of edges that need to update their higher-order support with little cost.

\subsection{The Improved Algorithm}
\label{sec:tia}

\algsetup{indent=0.5em}
\begin{algorithm}[t]
\caption{ $\kw{HOTDecom^+}(G, \tau)$}
{
%\small
\begin{algorithmic}[1]
\label{alg:improvebu}
%\FOR{\textbf{each} $v \in V(G)$, $e \in E(G)$}
%\STATE compute $d_\tau(v, G)$, $\underline{\phi}_{\tau}(e, G)$;
\FOR{\textbf{each} $e \in E(G)$}
\STATE compute $\underline{\phi}_{\tau}(e, G)$;

\ENDFOR
\STATE $k \leftarrow min_{e\in E(G)}(\underline{\phi}_{\tau}(e, G))$; 
%$k_{\kw{max}} \leftarrow \kw{max}_{v \in V(G)}\{d_\tau(v, G)\} + 1$;
%$k \leq k_{\kw{max}}$ \kw{and} 
\WHILE{$G\neq \emptyset$}
\FOR{\textbf{each} $e \in E(G)$ with $\underline{\phi}_\tau(e, G) = k$}
\STATE compute $\kw{sup}_\tau(e, G)$;
\ENDFOR
\WHILE{$\exists ~ e = (u, v)\in E(G)$ with $\kw{sup}_\tau(e, G) \leq k-2$}
\STATE $\phi_\tau(e, G) \leftarrow k$; $G' \leftarrow G$; $G\leftarrow G\setminus e$;

\FOR{\textbf{each} $e' = (u',v')\in E(G)$ with $u', v'\in \{u\} \cup \{v\}\cup \Delta_\tau(e, G') $}
%\IF {$\phi_\tau(e',G)$ has been obtained}
%\STATE \textbf{continue};
%\ENDIF
\IF {$\underline{\phi}_\tau(e',G')>k$ }
\STATE \textbf{continue}; $\quad\quad\quad\quad\quad$// \emph{delayed update}
\ENDIF
\IF {{ $\kw{pruneVertex}(G, u', k)$ or $\kw{pruneVertex}(G, v', k)$}}
\STATE \textbf{continue}; $\quad\quad\quad\quad\quad$ // \emph{early pruning}
\ENDIF
\STATE compute $\kw{dis}_G(u, w)$, $\kw{dis}_{G'}(u, w)$, $\kw{dis}_G(v, w)$, $\kw{dis}_{G'}(v, w)$ with $w \in \Delta_\tau(e, G')$;
\IF{$\kw{dis}_{G'}(u, u')$ = $\kw{dis}_{G}(u, u') \wedge \kw{dis}_{G'}(u, v')$ = $\kw{dis}_{G}(u, v') \wedge \kw{dis}_{G'}(v, u') $=$ \kw{dis}_{G}(v, u') \wedge \kw{dis}_{G'}(v, v')$=$\kw{dis}_{G}(v, v')$}
\STATE \textbf{continue}; $\quad\quad\quad\quad\quad$ // \emph{unchanged support detection}
\ENDIF
\IF{$\kw{sup}_\tau(e',G') > k-2$}
\STATE compute $\kw{sup}_\tau(e',G)$;
\ENDIF
\ENDFOR
\ENDWHILE
\STATE $k \leftarrow k+1$;
\ENDWHILE
\end{algorithmic}
}
\end{algorithm}

\algsetup{indent=0.5em}
\begin{algorithm}[t]
\caption{\kw{pruneVertex}($G,  v, k$)}
{
%\small
\begin{algorithmic}[1]
\label{alg:improvebupro}
\STATE $\mathcal{S} \leftarrow \emptyset$; compute $d_\tau(v,G)$;
\IF {$d_\tau(v, G)\le k-1$}
\STATE $ \mathcal{S}.\kw{put}(v)$;
\WHILE{$\mathcal{S}\ne\emptyset$}
\STATE $u \leftarrow \mathcal{S}.\kw{get}()$;
\FOR{\textbf{each} $e = (u, w)\in E(G)$ with $w \in N_1(u, G)$}
\STATE $\phi_\tau(e, G) \leftarrow k$;
\ENDFOR
\STATE $G' \leftarrow G$; $G\leftarrow G\setminus u$; $\mathcal{N} \leftarrow N_{\tau}(u, G')$;
\FOR{\textbf{each} $w \in \mathcal{N}$}
\STATE \textbf{if} $d_\tau(w, G)\le k-1$ \textbf{then} $ \mathcal{S}.\kw{put}(w)$; $\mathcal{N}\leftarrow \mathcal{N}\setminus w$;
\ENDFOR
\FOR{\textbf{each} $e'' = (x, y)\in E(G)$ with $x, y\in \mathcal{N}$ }
\STATE \textbf{if} $\underline{\phi}_\tau(e'', G') \leq k$ \textbf{then} compute $\kw{sup}_\tau(e'',G)$;
\ENDFOR
\ENDWHILE
\STATE \textbf{return} true;
\ELSE
\STATE \textbf{return} false;
\ENDIF
\end{algorithmic}
}
\end{algorithm}

With the delayed update strategy, early pruning strategy and unchanged support detection strategy, we are ready to present our improved algorithm \baselinep, which is shown in \refalg{improvebu}. 

\stitle{Algorithm.} \refalg{improvebu} shares a similar framework as \refalg{bu}. It first computes %$d_\tau(v, G)$ for each vertex   and
the lower bound $\underline{\phi}_\tau(e, G)$ for each edge (line 1-2). Then,  it decomposes  the graph in the increasing order of $k$. %It uses $k_\kw{max}$ to indicate the possible value of the maximum higher-order truss number (line 3). Here, $k_\kw{max}$ is the maximum value of $d_\tau(v, G)$ plus 1. 
 It first initializes $k$ as the minimum value of $\underline{\phi}_\tau(e, G)$ of all edges in $G$. 
 For a specific $k$, \refalg{improvebu}  first computes $\kw{sup}_{\tau}(e, G)$ for each edge with $\underline{\phi}_{\tau}(e, G) = k$ (line 5-6). Then, for the edge with $\kw{sup}_\tau(e, G) \leq k -2$, it assigns the higher-order truss number to  $e$ and removes $e$ from $G$ similarly as \refalg{bu} (line 7-8).  After removing $e$, \refalg{improvebu} updates the higher-order  support for the edges $e' = (u', v')$ incident to vertices in $\{u\} \cup \{v\} \cup \Delta_{\tau}(e, G')$. It first checks whether $\underline{\phi}_\tau(e', G') > k$. If it is true, \refalg{improvebu} delays the computation of $\kw{sup}_\tau(e', G)$ (line 10-11). Otherwise,  \refalg{improvebu} checks whether vertex $u'$ or $v'$ can be early pruned by  \kw{pruneVertex} (line 12-13). If these two vertices can not be pruned, it further checks whether the distance between $u$ (resp. $v$) and $u'$ (resp. $v'$) is changed  (line 15-16). If the distance is changed and $\kw{sup}_\tau(e', G') > k -2$ ,  \refalg{improvebu} recomputes $\kw{sup}_\tau(e', G)$ (line 18).  \refalg{improvebu} continues the above procedure until $G$ is empty (line 4).

\kw{pruneVertex}  prunes the vertices based on \reflem{prun}. It uses a set $\mathcal{S}$ to record the vertices that can be pruned. For a vertex $v$, it first computes $d_\tau(v, G)$ (line 1). If $d_\tau(v, G)$ is not greater than $k -1$, $v$ is put in $\mathcal{S}$ (line 2-3). Then,  it iteratively gets vertex $u$ from $\mathcal{S}$ until $\mathcal{S}$ is empty. For $u$, since $u$ can be pruned from $G$, which means the higher-order truss number for edges $e$ incident to $u$ is $k$, it assigns $k$ to $\phi_{\tau}(e, G)$ and removes the vertex from $G$ (line 6-8).       Due to the removal of $u$, the $\tau$-hop degree for the vertices in $N_{\tau}(u, G')$ may be changed. Therefore, it further checks the $\tau$-hop degree of these vertices. If these vertices  can be pruned, then  they are put in $\mathcal{S}$ and removed from $\mathcal{N}$ (line 8-10). For the edges whose two incident vertices are in the  $\mathcal{N}$, it recomputes their higher-order edge support if $\underline{\phi}_\tau(e'', G') \leq k$ (line 11-12). If the procedure prunes any vertices, it returns true (line 13). Otherwise, it returns false (line 15).

\begin{figure}[t]
\begin{center}
\includegraphics[width=0.85\columnwidth]{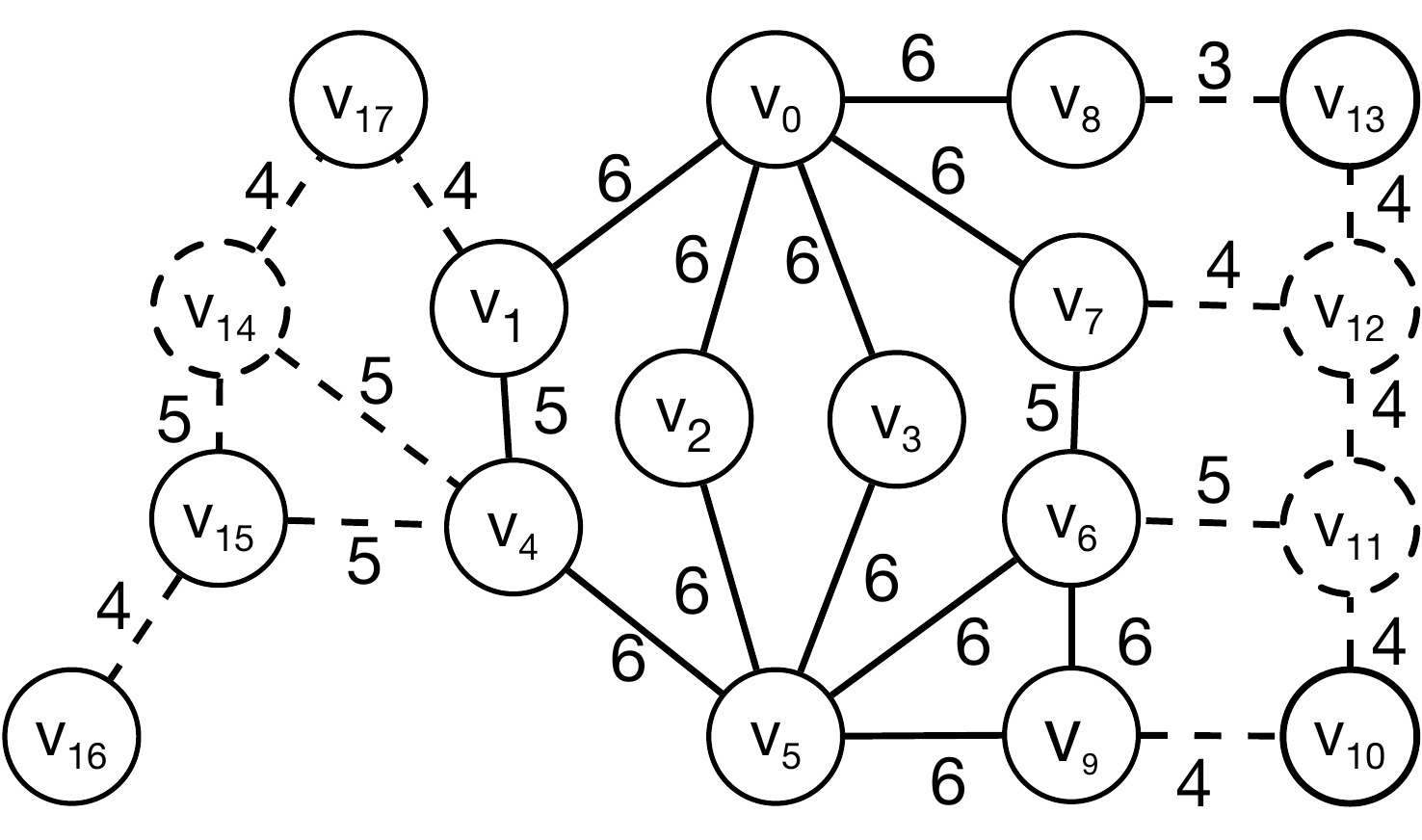}
\end{center}
%\vspace{-0.4cm}
\caption{The procedure of $\kw{HOTDecom^+}$}
\label{fig:example_5}
%\vspace{-0.4cm}
\end{figure}

\begin{example}
Reconsider the graph $G$ shown in \reffig{example_2}. \reffig{example_5} shows the procedure  of \baselinep. It first computes lower bound $\underline{\phi}_2(e, G)$ for each edge. $\underline{\phi}_2(e, G)$ is shown near each edge in \reffig{example_2}.  After that,  it  starts the decomposition with $k = 3$. Since $\underline{\phi}_2((v_{8}, v_{13}), G)=3$, it computes $\kw{sup}_2((v_{8}, v_{13}), G) = 3>k-2$. Then, $k$ becomes 4 and  it computes $\kw{sup}_2(e, G)$ for edges with $\underline{\phi}_2(e, G)=4$. It  gets $\kw{sup}_2((v_{15},v_{16}),G)=2$, $\kw{sup}_2((v_{14},v_{15}),G)=5$, $\dots$. Then, it removes $(v_{15},v_{16})$. Due to the removal of $(v_{15},v_{16})$ , $\kw{sup}_2((v_{14}, v_{15}), G)$ is updated from 5 to 4, since $\underline{\phi}_2((v_{4}, v_{14}), G)=5$$>$$4$ and $\underline{\phi}_2((v_{4}$,$v_{15})$,$G)=5$$>$$4$, it does not  update their higher-order support. Then, $k$ becomes 5, since $\kw{sup}_2((v_{14},v_{17}),G)=3\leq 5-2$, $(v_{14},v_{17})$ is removed. Here, although $v_1, v_4 \in \Delta_2((v_{14}, v_{17}), G)$, it does not update the higher-order support of $(v_{1},v_{4})$, $(v_{1},v_{17})$ and $(v_{4},v_{17})$ since the distance between $v_1$ (resp. $v_4$) and $v_{14}$ (resp. $v_{17}$) is unchanged after the removal of $(v_{14}, v_{17})$. Meanwhile, due to the removal of $(v_{14},v_{17})$, $d_2(v_{14}, G)$ is updated from 5 to 4, hence $v_{14}$ and its incident edges are removed  directly. $(v_{4},v_{15})$ is removed following $v_{14}$. Here, although $v_4,v_5 \in \Delta_2((v_{4}, v_{14}) \cap \Delta_2((v_{4}, v_{15})$, it does not update the support of $(v_{4}, v_5)$ as $\underline{\phi}_2((v_{4}, v_{5}), G)>5$. Similarly, edges and vertices are removed  until $G$ is empty. As shown in this example, many higher-order support updates are avoided compared with \baseline shown in \refex{baseline}.
\end{example}

\begin{theorem}
\label{thm:improvebuc}
Given a graph $G$ and an integer $\tau$, \refalg{improvebu} computes the higher-order truss number for each edge  correctly.
\end{theorem}

\begin{proof}
Following \refthm{buc}, we only need to show that the reduced higher-order updates caused by line 11-16 of \refalg{improvebu} do not affect the correctness of the algorithm. This can be guaranteed by \reflem{tdiameter}, \reflem{edgediameter},  \reflem{vertexprune}, and \reflem{unchanged}.
\end{proof}

\begin{theorem}
\label{thm:c}
The time complexity of \refalg{improvebu} is $O(m\cdot m_\tau \cdot (m_\tau+n_\tau))$.
\end{theorem}

Although \baselinep shares the same worst case time complexity as \baseline, lots of  higher-order edge support updates can be reduced in practice as verified in our experiments, which makes \baselinep significantly outperforms \baseline.

\section{Top-$r$ Higher-Order Trusses Computation}
\label{sec:td}

In the above section, we investigate the higher-order truss decomposition problem that  the $(k, \tau)$-trusses with all possible $k$  values are computed. However, in some applications, users  are only interested in the $(k, \tau)$-truss with a large $k$ value since the $(k, \tau)$-truss with  large $k$ value is generally more cohesive and represents the core part of the graph \cite{DBLP:series/ads/LeeRJA10,DBLP:journals/pvldb/YuanQLCZ16}.  Certainly, we can use \baselinep directly to address this problem. Nevertheless, as this approach adopts the bottom-up decomposition paradigm, the $(k, \tau)$-trusses with small $k$ values have to be computed, which leads to lots of unnecessary computation.   To address this problem, in this section, we propose a new approach tailored for the top $r$ $(k, \tau)$-trusses computation problem. Formally, given a graph $G$ and two integers $r$ and $\tau$, the top $r$ $(k, \tau)$-trusses  computation problem returns the $(k, \tau)$-trusses with $k$ values in the range of $(k_\kw{max}-r , k_{\kw{max}}]$,  where $k_\kw{max}$ is the maximum value of $k$ such that there is a non-empty $(k, \tau)$-truss regarding the corresponding $k$ value in the graph.

\stitle{An upper-bound integrated approach.} To conduct the top $r$ $(k, \tau)$-trusses computation, suppose that we have known the $\phi_{\tau}(e, G)$ for each edge $e$ in $G$ in prior, we can take the edges $\Phi_{\tau, \phi > k_\kw{max}-r}$  as the input of \baselinep and conduct the decomposition starting from $k_{\kw{max}} -r + 1$ to compute the result. In this way, the unnecessary computation involved in  \baselinep can be totally avoided. However, this approach has to know  $\phi_{\tau}(e, G)$ for each edge $e$ in $G$ in prior, which is  intractable. On the other hand, if we can obtain an upper bound $\overline{\phi}_{\tau}(e, G)$ for each edge $e$ in $G$, we can take the edges with $\overline{\phi}_{\tau}(e, G) > k_{\kw{max}} - r$ as the input of \baselinep and obtain the correct answer for the similar reason as \baselinep.  Following this idea, we propose an upper bound integrated approach to realize the top $r$ $(k, \tau)$-trusses computation. We first present the upper bound of $\phi_\tau(e, G)$.

\stitle{Upper bound of $\phi_\tau(e, G)$.} For an  edge $e$ in a given graph $G$, a direct upper bound of $\phi_\tau(e, G)$ is $\kw{sup}_\tau(e, G)$. However, this bound is too loose.  To obtain a tight upper bound, we define:

\begin{definition}
\label{def:ubsubgraph}
\textbf{(Support-Vertex Bounded Subgraph)} Given a graph $G$ and an integer $\tau$, for an edge $e = (u, v) \in E(G)$, let $G'$ be the subgraph induced by $\{u\} \cup \{v\} \cup \{\Delta_{\tau}(e, G)\}$, the  support-vertex bounded subgraph of $e$, denoted by $\Lambda(e)$, is a connected  subgraph $G''$ of $G'$ such that (1) $e \in E(G'')$ (2) for each edge $e \in E(G'')$, $\kw{sup}_{\tau}(e, G) \geq |V(G'')|-2$ (3) $|V(G'')|$ is maximal.
\end{definition}

\begin{lemma}
Given a graph $G$ and an integer $\tau$, for an edge $e = (u, v)$, $\phi_{\tau}(e, G) \leq |V(\Lambda(e))|$.
\end{lemma}

\begin{proof}
We can prove it by contraction. Assume that there exists an edge $e' = (u', v')$ such that $\phi_{\tau}(e', G) > |V(\Lambda(e'))|$. Based on \refdef{trussnum}, let $\mathbb{G}$ be the subgraph of $G''$ induced by $\{u'\} \cup \{v'\} \cup \{\Delta_{\tau}(e', G)\}$, where $G''$ is the $(\phi_{\tau}(e', G), \tau)$-truss in $G$ containing $e'$. Based on \refdef{truss}, for each edge $e'' \in E(\mathbb{G})$, $\kw{sup}_{\tau}(e'', G'') \geq \phi_{\tau}(e'', G)-2$. Moreover, $\kw{sup}_{\tau}(e'', G) \geq \kw{sup}_{\tau}(e'', G'')$. Therefore, $\kw{sup}_{\tau}(e'', G) \geq \phi_{\tau}(e'', G)-2 \geq \phi_{\tau}(e', G)-2 > |V(\Lambda(e'))| - 2$. Meanwhile, $|V(\mathbb{G})| \geq \phi_{\tau}(e', G) > |V(\Lambda(e'))|$. It means $\mathbb{G}$ satisfies condition (1) and (2) of \refdef{ubsubgraph} but the number of vertices is bigger than $|V(\Lambda(e'))|$, which contracts with condition (3) of \refdef{ubsubgraph}. Thus, the lemma holds.
\end{proof}

\begin{definition}
\label{def:ub}
\textbf{(Upper Bound $\overline{\phi}_{\tau}(e, G)$)} Given a graph $G$ and an integer $\tau$, for an edge $e  \in E(G)$, $\overline{\phi}_{\tau}(e, G) = |V(\Lambda(e))|$.
\end{definition}

\algsetup{indent=1em}
\begin{algorithm}[t]
\caption{ $\kw{HOTTopR}(G, \tau, r)$}
{
%\small
\begin{algorithmic}[1]
\label{alg:td}
\FOR{\textbf{each} $e \in E(G)$}
\STATE  compute $\underline{\phi}_{\tau}(e, G)$; $\overline{\phi}_{\tau}(e, G)\leftarrow \kw{computeUB}(e, G)$;
\ENDFOR
\STATE  $k_{\kw{max}} \leftarrow \kw{max}_{ e \in E(G)}\{\overline{\phi}_{\tau}(e, G)\}$;  $\mathcal{G} \leftarrow \emptyset$; First $\leftarrow true$;
%$ k  \leftarrow k_{\kw{max}}-r+1$;
%\FOR {\textbf{each} $e \in E(G)$ with $\overline{\phi}_{\tau}(e, G) > k_{\kw{max}} - r $}
%\STATE $\mathcal{G} \leftarrow \mathcal{G} \cup e$;
%\ENDFOR

%\WHILE{$k\leq k_{max}$}
%\STATE line 4-19 of \refalg{improvebu} replacing $G$ with $\mathcal{G}$;
%\ENDWHILE

\WHILE{$\Phi_{\tau, \phi = k_{\kw{max}}} = \emptyset$}
\IF{!First}
\IF{$\exists ~e \in E(\mathcal{G}) $ with $\phi_\tau(e, G)$ obtained}
\STATE $k_{\kw{max}} \leftarrow \kw{max}_{ e \in E(\mathcal{G})}\{\phi_{\tau}(e, G)\}$;
\ELSE
\STATE $k_{\kw{max}} \leftarrow k_{\kw{max}} - r$;
\ENDIF
\ENDIF
\STATE $ k  \leftarrow k_{\kw{max}}-r +1$;
\STATE First $\leftarrow false$;
\FOR {\textbf{each} $e \in E(G)$ with $\overline{\phi}_{\tau}(e, G) > k_{\kw{max}} - r  $}
\STATE $\mathcal{G} \leftarrow \mathcal{G} \cup e$;
\ENDFOR
\FOR{\textbf{each} $e \in E(\mathcal{G})$}
\IF{ $\phi_\tau(e, G)$ has been obtained}
\STATE $\kw{sup}_\tau(e, \mathcal{G}) \leftarrow \phi_\tau(e, G)-2$;
\ELSE
\STATE compute $\kw{sup}_\tau(e, \mathcal{G})$;
\ENDIF
\ENDFOR

\WHILE{$k \leq k_{\kw{max}}$}
\STATE line 7-10 of \refalg{improvebu} replacing $G$ with $\mathcal{G}$;
\IF {$\phi_\tau(e',G)$ has been obtained}
\STATE \textbf{continue};
\ENDIF
\STATE line 11-19 of \refalg{improvebu} replacing $G$ with $\mathcal{G}$ ;
\ENDWHILE
\ENDWHILE
\vspace{0.2cm}
\STATE \textbf{Procedure} \kwnospace{computeUB}$(e = (u, v), G)$
\STATE $l \leftarrow 2$; $r \leftarrow \kw{sup}_\tau(e, G)+2$;
\WHILE{$l \leq r$}
\STATE $\kw{mid} \leftarrow (l+r)/2$;
\STATE construct $G''$ by $\tau$-hop BFS traversal starting from $u$ and $v$ through edges with $\kw{sup}_{\tau}(e, G) \geq \kw{mid}-2$;
\IF{$|V(G'')| <  \kw{mid}$}
\STATE $r \leftarrow \kw{mid} -1$;
\ELSE
\STATE $\overline{\phi} \leftarrow \kw{mid}$;  $l \leftarrow \kw{mid} + 1$;
\ENDIF
\ENDWHILE
\STATE \textbf{return } $\overline{\phi}$ ;
\end{algorithmic}
}
\end{algorithm}

\stitle{Algorithm.} With the  upper bound, our top $r$ $(k, \tau)$-trusses computation algorithm \baselines is shown in \refalg{td}. It first computes  $\underline{\phi}_{\tau}(e, G)$ and $\overline{\phi}_{\tau}(e, G)$ for each edge in $G$ (line 1-2). Then,  it  retrieves the maximum value of $\overline{\phi}_{\tau}(e, G)$ among all the edges (line 3). After that, the graph $\mathcal{G}$ consisting the edges with $\overline{\phi}_{\tau}(e, G) > k_{\kw{max}} -r $  is constructed (line 12-13). Then, \refalg{td} computes the top $r$ $(k, \tau)$-trusses by utilizing \refalg{improvebu}. Since \refalg{td} uses the maximum value of $\overline{\phi}_{\tau}(e, G)$ among all the edges as $k_{\kw{max}}$ in line 3, it is possible that there exists no such $(k_{\kw{max}}, \tau)$-truss in $G$. In this case, \refalg{td} further explores the possible $k_{\kw{max}}$ of $G$. It considers two subcases: (1) there exist some edges whose higher-order truss number have been obtained. In this case, it can be easily derived that $k_{\kw{max}}$ is the maximum value of $\phi_{\tau}(e, G)$ among these edges (line 6-7). (2) there exists no edge whose higher-order truss number is in the range of $(k_{\kw{max}}-r,k_{\kw{max}}]$. In this case, \refalg{td} progressively decreases the value of $ k_{\kw{max}}$ by $r$ until $k_{\kw{max}}$ is obtained (line 9).  After that, it continuously search search the result with new value of $k_{\kw{max}}$ and $k$. Note that for the edges  $\phi_\tau(e, G)$ have been obtained, it does not need to recompute $\kw{sup}_\tau(e, \mathcal{G})$ and just sets $\kw{sup}_\tau(e, \mathcal{G})$ as $\phi_\tau(e, G)-2$, which can reduce the unnecessary computation (line 15-16). Similarly, the value of $\kw{sup}_\tau(e, \mathcal{G})$ for these edges is not necessary to be updated in line 21-22. The algorithm terminates when top $r$ $(k,\tau)$-trusses are found.

Procedure \kwnospace{computeUB} is used to compute $\overline{\phi}_{\tau}(e, G)$ for an edge $e$. It adopts a binary search strategy to find $\Lambda(e)$ and uses $l$ and $r$ to indicates the current search range. In each iteration, it starts two $\tau$-hop BFS traversal from $u$ and $v$ respectively by visiting the edges with $\kw{sup}_{\tau}(e, G) \geq \kw{mid} -2$ to construct $G''$ (line 28). If there exists $G''$ such that $|V(G'')| \geq \kw{mid}$, $\overline{\phi}$ records current possible upper bound of $\phi_\tau(e, G)$ (line 32). The search continues until $l > r$ and returns $\overline{\phi}$.

\begin{theorem}
Given a graph $G$, an integer $\tau$ and an integer $r$, \refalg{td} computes the top $r$ $(k, \tau)$-trusses in $G$ correctly.
\end{theorem}

\myproof
This theorem can be proved similarly as \refthm{improvebuc}.
\eop

\begin{theorem}
\label{thm:tdtime}
The time complexity of \refalg{td} is $O( r' \cdot m'\cdot m_\tau \cdot (m_\tau+n_\tau) + m \cdot \log n_\tau \cdot (m_\tau+n_\tau))$, where $r'$ is the number of iterations in line 4, $m'$ is the maximum number of edges of $\mathcal{G}$.
\end{theorem}

\myproof
This theorem can be proved similarly as \refthm{butime}.
\eop

Although the time complexity  of \refalg{td} is not reduced compared with \refalg{improvebu} theoretically,  \refalg{td} is efficient in terms of the top $r$ $(k, \tau)$-trusses computation in practice. This is because the top $r$ $(k, \tau)$-trusses are generally much smaller than the input graph and the proposed upper bound shown in \refdef{ub} is effective.

\section{PERFORMANCE STUDIES}
\label{sec:exp}

In this section, we present our experimental results. All the experiments are conduct on a machine with 4 Intel Xeon 3.0GHz CPUs and 64GB RAM running Linux.% ( Red Hat Linux 4.8, 64 bit).

\begin{table}[h]
\topcaption{Statistic of the real datasets}
\vspace{0.2cm}
\label{tab:dataset}
\centering
\def\arraystretch{1.0}
\setlength{\tabcolsep}{0.50em}
{\small
\begin{tabularx}{\linewidth}{p{0.15\linewidth}<{\centering}|p{0.2\linewidth}<{\centering}|p{0.15\linewidth}<{\raggedleft}|p{0.145\linewidth}<{\raggedleft}|p{0.14\linewidth}<{\raggedleft}}

% \begin{tabularx}{\linewidth}{@{\extracolsep{\fill}}c|c|c|c|c}
\hline
\cg Datasets &\cg Type & \cg $|V|$ & \cg $|E|$ &\cg $d_{max}$ \\
\hline
\kw{PT}&Biography&1,870&2,203&56\\
 \kw{CG}&Collaboration &5,242&14,496&81\\
 %&43 \\
% 17
%\kw{Email}-\kw{EU}&
 \kw{EM}&E-Mail &1,005&25,571&345\\
 %&34\\
 %7
%\kw{Hams-Friend}&
%\kw{HF}&1,858&12,534&272&14 \\
%\kw{Ca-Hepth}&
\kw{CH}&Collaboration &9,877 &25,998&65\\
%&31\\
%17
%\kw{Ego-Facebook}&
\kw{FB}&Social &4,039&88,234&1,045\\
%&115 \\
%8
%\kw{Coma-Mazon}&
\kw{CA}&Product &334,863 &925,872&549\\
%&6\\
%44\\
%\kw{Com-Dblp}&
\kw{CD}&Collaboration &317,080&1,049,866&343\\
%&113\\
%21\\
\kw{AM}&Purchasing  &262,111&1,234,877&420\\
%&6\\
%&32\\
%\kw{web-NotreDame}&
\kw{WN}&Web &325,729&1,497,134&10,721\\
%&46\\
%\KW{Dbpeida-Team}&
%\kw{DT}&935,591&1,366,466&2,671&\\
%\kw{Web-Google}
\kw{WG}&Web &875,713&5,105,039&6,332\\
%&155\\
%21
%\kw{Dblp-author}&
\kw{DB}&Collaboration &4,000,150&8,649,005&954\\
%&11\\
%\kw{Cit-Patent}&
\kw{CP}&Citation &3,774,768&16,518,948&793\\
%&64\\
%22\\
%\kw{US}&3,774,768&16,518,947&793&22\\
\hline
\end{tabularx}
}
%\vspace{-0.4cm}
\end{table}

\begin{figure}[h]
\begin{center}
\begin{tabular}{c}
\subfigure[$\tau$=2]{
\includegraphics[width=0.85\columnwidth]{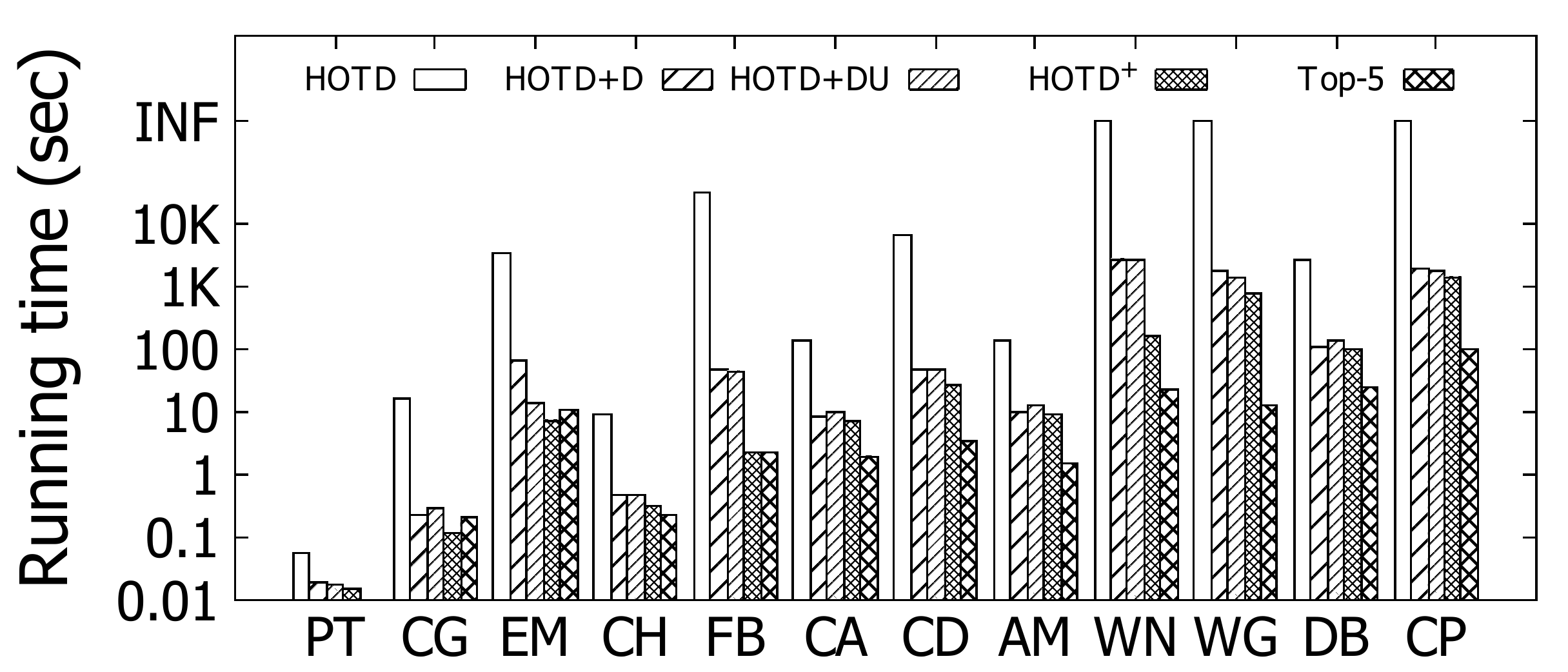}
\label{fig:time_2}}
\vspace{-0.2cm}
\\
\subfigure[$\tau$=3]{
\includegraphics[width=0.85\columnwidth]{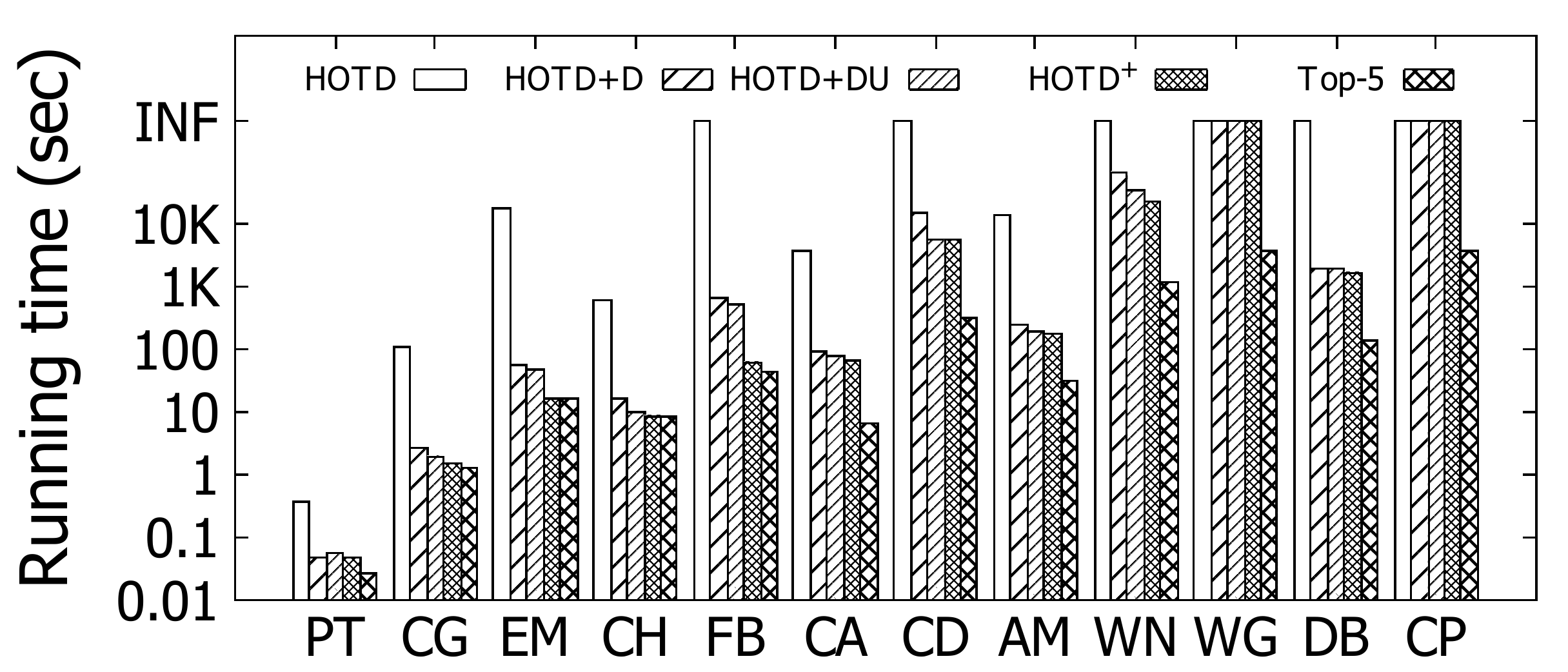}
\label{fig:time3}}
\vspace{-0.2cm}

\\
\subfigure[$\tau$=4]{
\includegraphics[width=0.85\columnwidth]{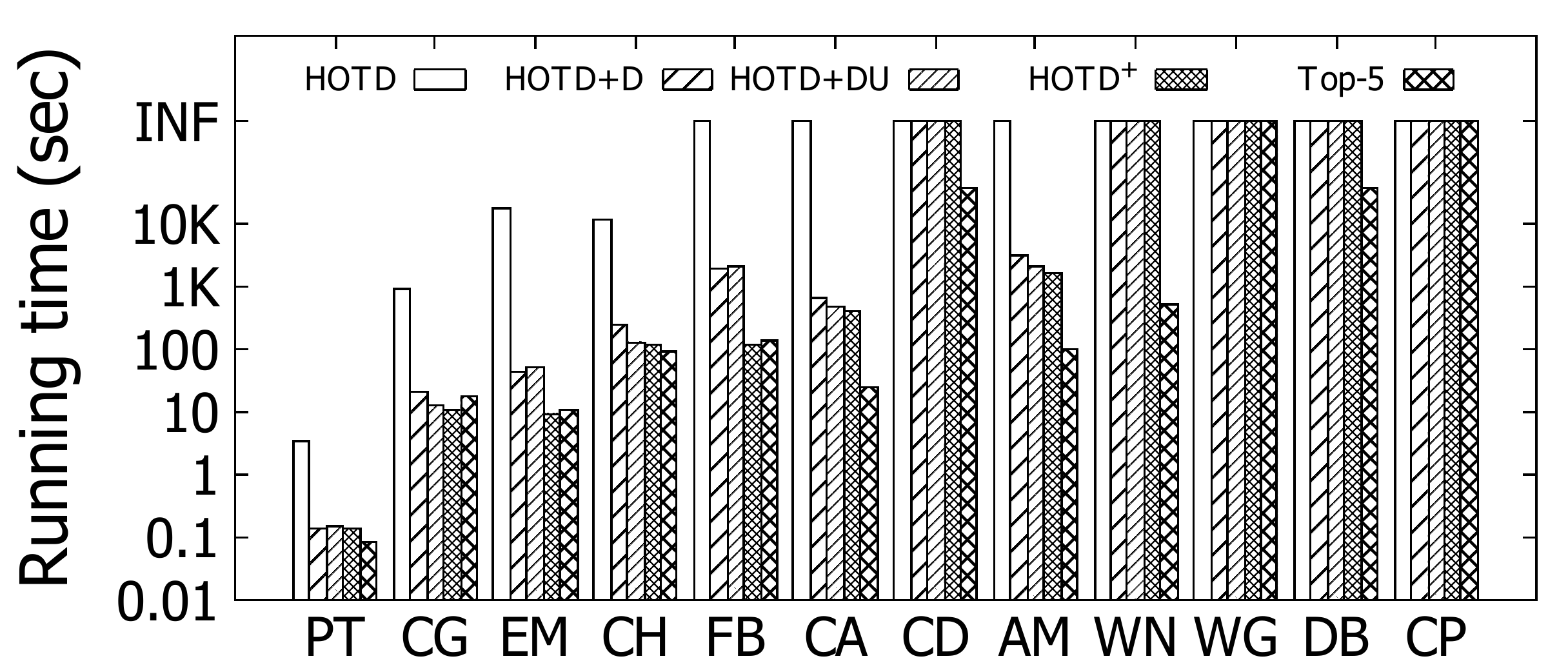}
\label{fig:time3}}

\end{tabular}
\end{center}
\vspace{-0.4cm}
\caption{The running time of algorithms}
\label{fig:time}
%\vspace{-0.4cm}
\end{figure}

 \begin{figure}[h]
\begin{center}
\begin{tabular}{c}
\subfigure[$\tau$=2]{
\includegraphics[width=0.85\columnwidth]{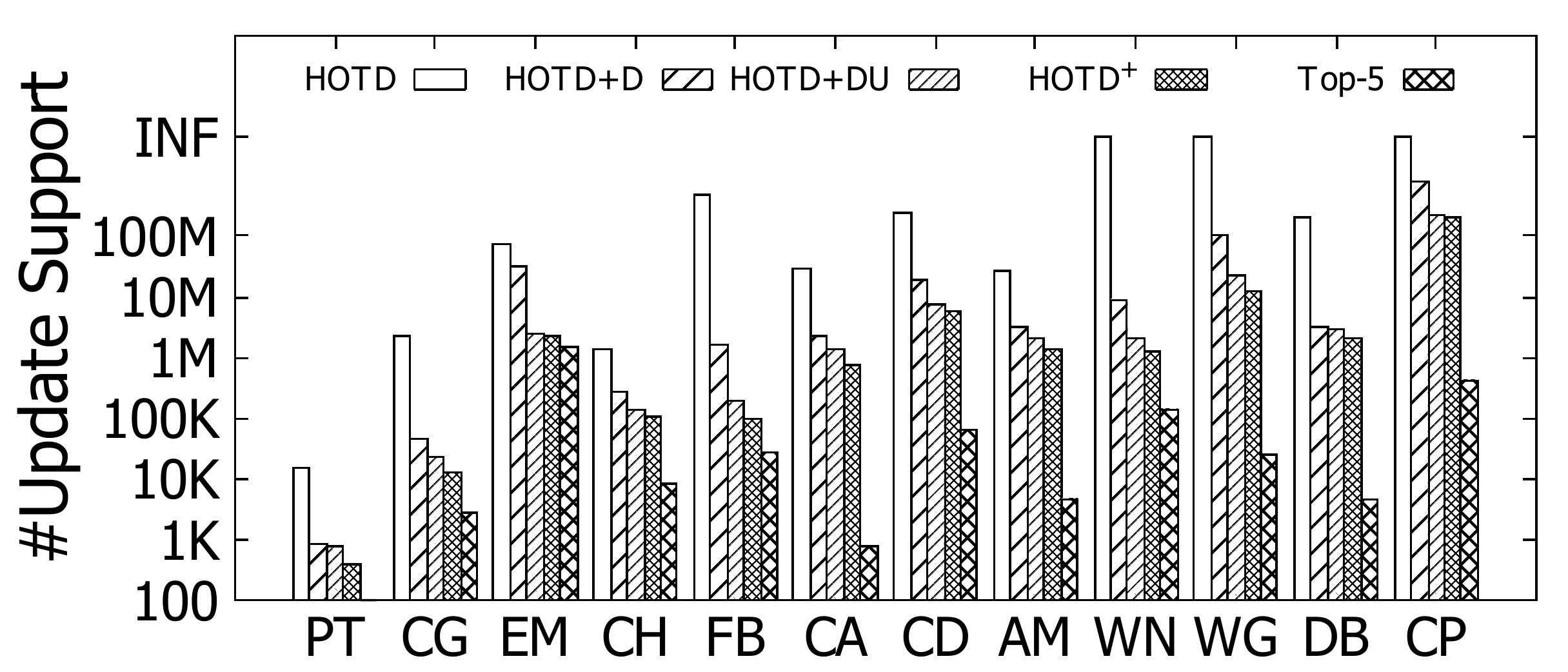}
\label{fig:time_2}}
\vspace{-0.2cm}
\\
\subfigure[$\tau$=3]{
\includegraphics[width=0.85\columnwidth]{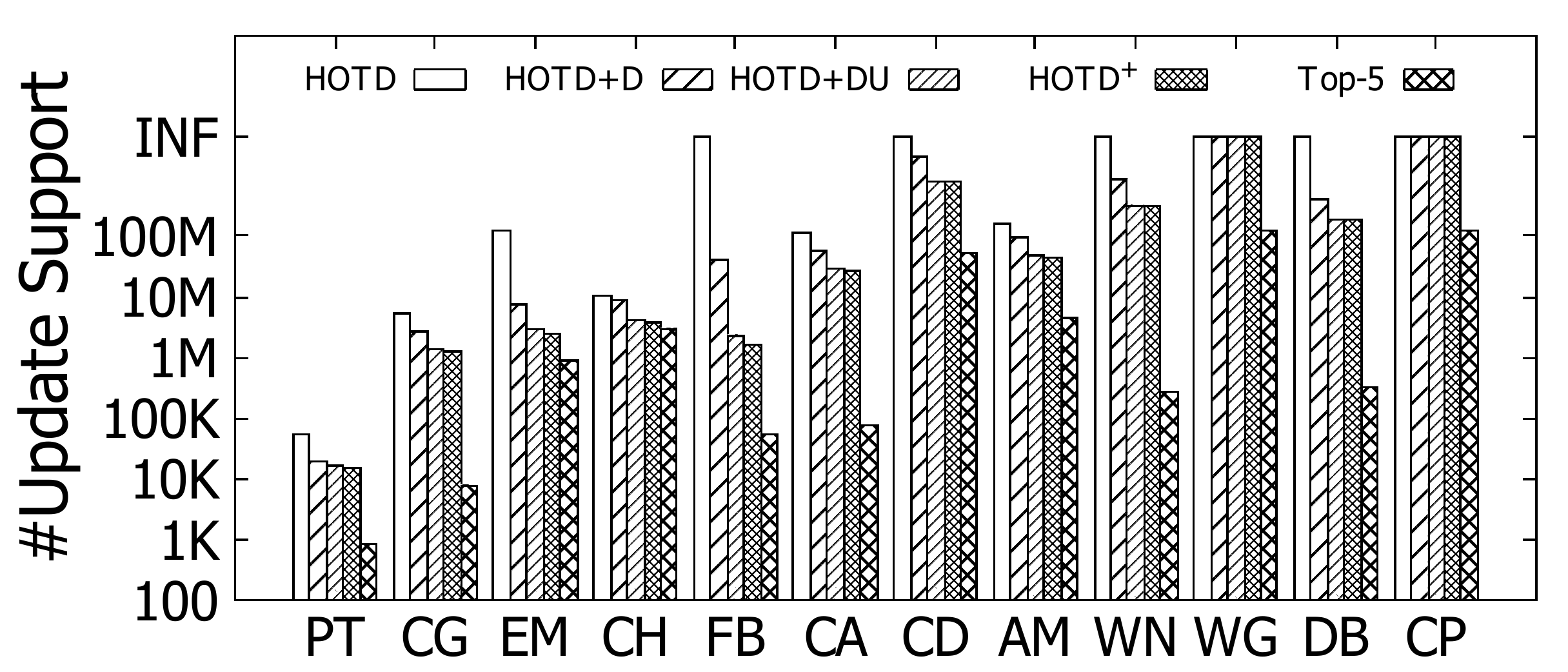}
\label{fig:time3}}
\vspace{-0.2cm}

\\
\subfigure[$\tau$=4]{
\includegraphics[width=0.85\columnwidth]{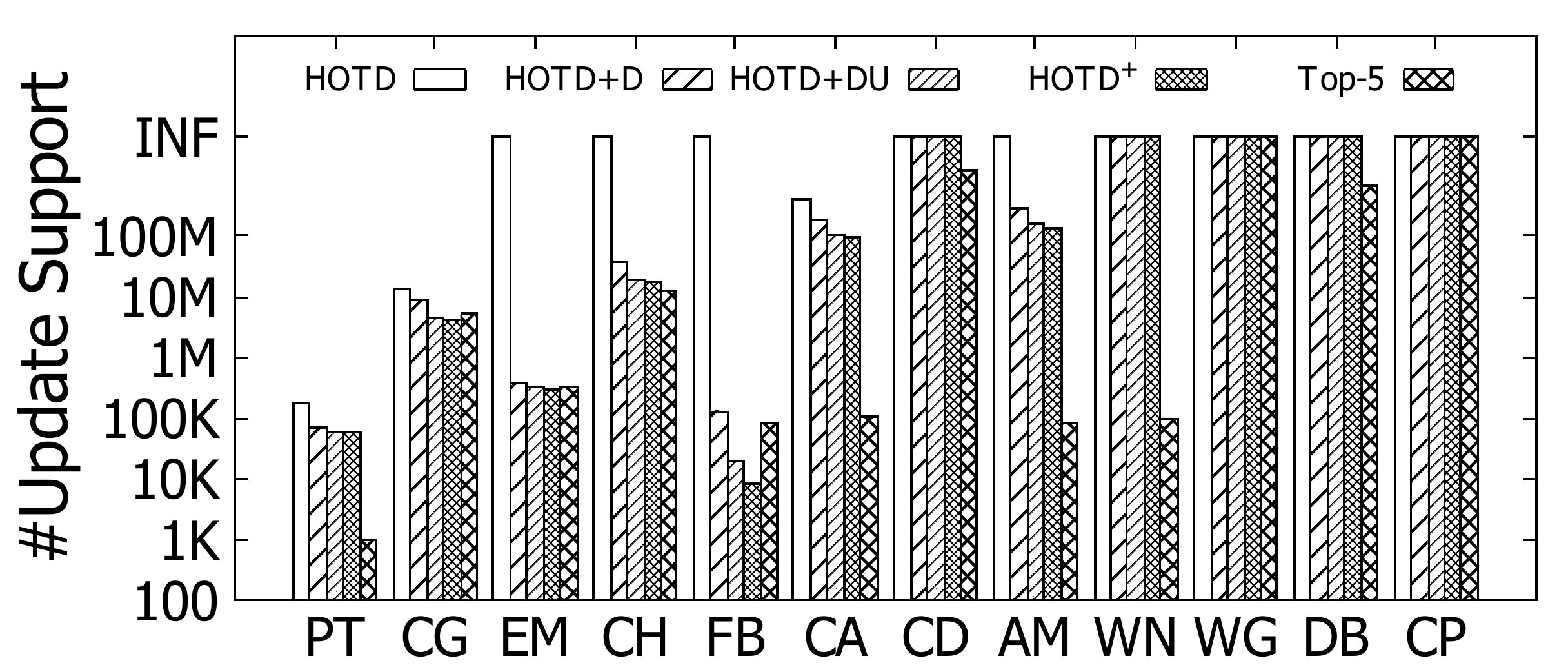}
\label{fig:time3}}

\end{tabular}
\end{center}
\vspace{-0.4cm}
\caption{The number of the support update}
\label{fig:updatecnt}
\vspace{-0.4cm}
\end{figure}
\vspace{-0.5cm}

\stitle{Datasets.} We evaluate our algorithms on 12 real world datasets. \kw{PT} is downloaded from KONECT\footnote{\url{http://konect.cc/}}, \kw{DB} is downloaded from Network Repository\footnote{\url{http://networkrepository.com/}} and the remaining datasets are downloaded from SNAP\footnote{\url{http://snap.stanford.edu/data/index.html}}. \reftable{dataset} shows the details of the datasets, where $d_{max}$ is the maximum degree of vertices in graphs.

%\footnote{For the reproducibility, our code is anonymously released at: \url{https://drive.google.com/file/d/134UZ6IgKRWSaguNe1vzEpBRy6RL2UykY/view?usp=sharing}}

\stitle{Algorithms.} We compare the following algorithms:\\
(1) \hotd: \baseline algorithm. \\(2) \hotdd: \baseline algorithm with the delayed update strategy. \\(3) \hotddu: \hotdd algorithm with the unchanged support detection strategy. \\(4)  \hotdp: \baselinep algorithm. \\(5) \kw{TOP}-5: \baselines algorithm with $r=5$.

All algorithms are implemented in C++, using g++ compiler with -O3.  Let $\tau$=[2-4], since under large $\tau$  ($\tau>4$), it will lead to the loss of cohesiveness of $(k, \tau)$-truss.
Thereby, users can choose $\tau$ from $[2-4]$ according to their requirement for cohesiveness of result. All reported results were averaged over 5 repeated runs. If an algorithm cannot finish in 12 hours, we denote the processing time as \kw{INF}. 
%If all algorithms cannot get the result on a dataset within 12 hours, we omit the running time of the dataset. 
%We first evaluate the efficiency of our proposed algorithms. Then, we evaluate the characterization of the $(k, \tau)$-truss model by comparing it with the traditional $k$-truss model. 

%\subsection{Efficiency Evaluation}

%In this part, we evaluate the efficiency of our proposed algorithms.

%\stitle{Exp-1: Efficiency of our proposed algorithms.} \reffig{time} shows the  running time of the four algorithms on 11 datasets when $\tau = 2, 3, 4$. 

\stitle{Exp-1: Efficiency of our proposed algorithms.} In this experiment, we compare the  running time of five algorithms on all datasets when $\tau$=2,3,4. The results are shown in \reffig{time}.

As $\tau$ increases, the running time of the algorithms increases as well. Regarding the algorithms, \hotd is slowest among the five algorithms.  \hotdd is much faster than \hotd benefited from the delayed update strategy. Moreover, with utilizing the unchanged support detection strategy, \hotddu is further faster than \hotdd. Afterwards, \hotdp is more efficient than \hotddu. This is because \hotdp adopts all three optimization strategies to reduce the number of edges that need to update their higher-order support. Compared with \hotd, \hotdp  achieves up to 4 orders of magnitude speedup. Besides, compared with \hotdp, \kw{HOTTopR} is suitable to compute the top $r$ results as it avoids lots of unnecessary computation related to the non-top-$r$ results in \hotdp.

%As $\tau$ increases, the running time of the algorithms increases as well. Regarding the algorithms, \hotd is slowest among the five algorithms.  \hotdd and \hotddu are much faster than \hotd.  \hotdp is more efficient than \hotdd and \hotddu. This is because \hotdp adopts three optimization strategies to reduce the number of edges that need to update their higher-order support. Compared with \hotd, \hotdp  achieves up to 4 orders of magnitude speedup. Compared with \hotdp, \kw{HOTTopR} is suitable to compute the top $r$ results as it avoids lots of unnecessary computation related to the non-top-$r$ results in \hotdp.

%\vspace{-0.1cm}
\stitle{Exp-2: Number of higher-order  support update.} In this experiment, we report the number of higher-order support update of all algorithms during the decomposition when $\tau = 2,3,4$. The results are shown in \reffig{updatecnt}.

As shown in \reffig{updatecnt}, on most datasets, the number of higher-order support updates of \hotd increase when $\tau$ increases from $2$ to $4$. This is because as the value of $\tau$ increases, the scope of edges that need update enlarges as well. However, the number of higher-order support update is significantly reduced by the optimization strategies. The results also explain the reasons causing different running times of the algorithms shown in \reffig{time}.

\begin{table}[h]
\topcaption{Tightness of $\underline{\phi}_\tau(e, G)$}
\vspace{0.2cm}
\label{tab:lbacc}
\centering
\def\arraystretch{1.1}
\setlength{\tabcolsep}{0.50em}
{\small
\begin{tabular*}{\linewidth}{p{0.11\linewidth}<{\centering}|p{0.075\linewidth}<{\centering}|p{0.075\linewidth}<{\centering}|p{0.075\linewidth}<{\centering}|p{0.11\linewidth}<{\centering}|p{0.074\linewidth}<{\centering}|p{0.074\linewidth}<{\centering}|p{0.074\linewidth}<{\centering}}
\specialrule{0em}{0.5pt}{0pt}
%     \cline{2-5}
     \hline
 \cg Dataset&\cg $\tau$=2 &\cg  $\tau$=3 &\cg $\tau$=4  &\cg Dataset&\cg $\tau$=2 & \cg $\tau$=3 &\cg $\tau$=4 \\
\hline
\kw{PT}&0.02&0.16&0.20 &\kw{CD}&0.06&0.60&-\\
\kw{CG}&0.03&0.36&0.31 &\kw{AM}&0.11&0.39&0.36\\

\kw{EM}&0.34&0.49&0.03 &\kw{WN}&0.03&0.09&-\\
%\kw{HF}&0.30&0.60&0.17\\
\kw{CH}&0.07&0.54&0.47 &\kw{WG}&0.06&-&-\\
\kw{FB}&0.002&0.06&0.10 &\kw{DB}&0.01&0.15&-\\
\kw{CA}&0.07&0.27&0.28 &\kw{CP}&0.15&-&-\\
\hline
\end{tabular*}
}
\end{table}

\vspace{-0.5cm}

\begin{figure}[h]
\begin{center}
\begin{tabular}{c}
\subfigure[\kw{Random} \kw{graphs}  ($\tau$=2)]{
%\hspace{-0.5cm}
\includegraphics[width=0.85\columnwidth]{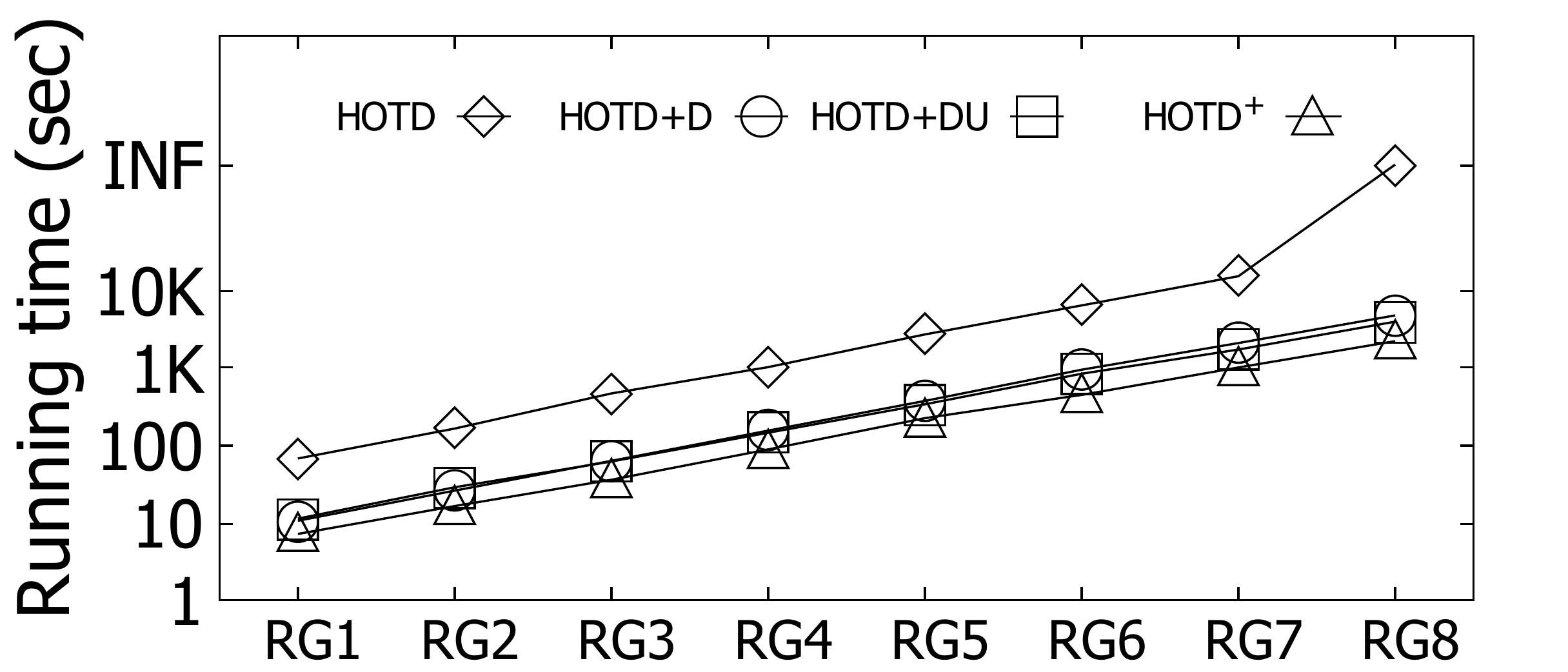}
\label{fig:scal_rg2}}
\vspace{-0.2cm}

\\
%\vspace{-0.2cm}
\subfigure[\kw{Random} \kw{graphs} ($\tau$=3)]{
\includegraphics[width=0.85\columnwidth]{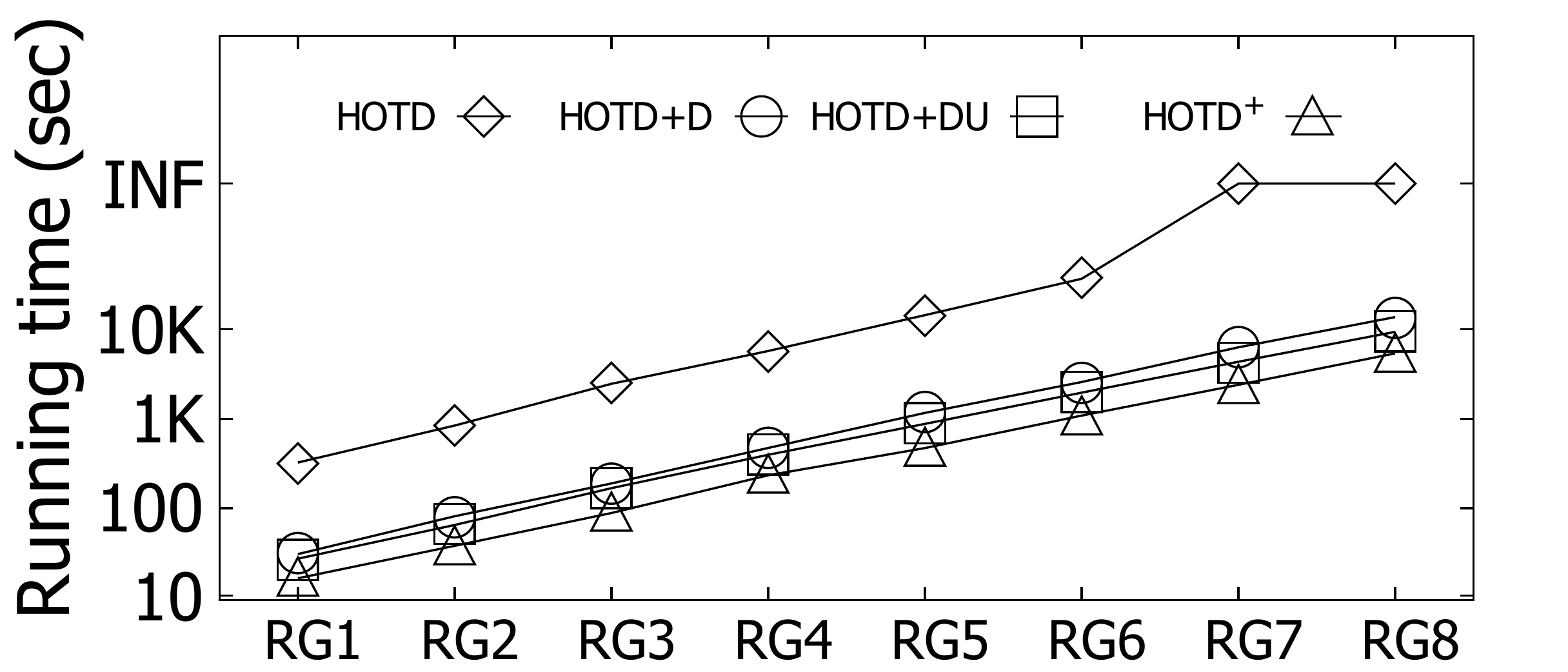}
\label{fig:scal_rg3}}
\vspace{-0.2cm}

\\
\subfigure[\kw{Random} \kw{graphs} ($\tau$=4)]{
%\hspace{-0.5cm}
\includegraphics[width=0.85\columnwidth]{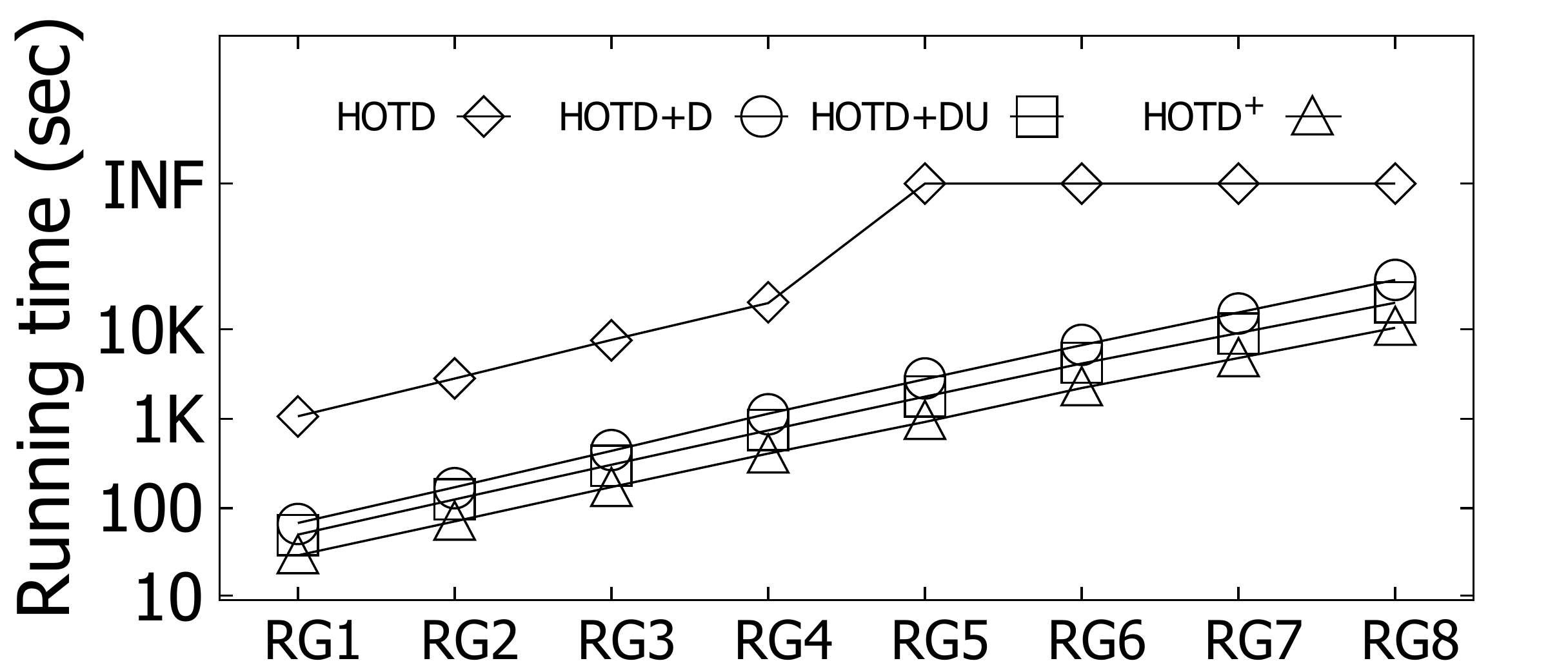}
\label{fig:scal_rg4}}
%\vspace{-0.2cm}
\end{tabular}
\end{center}
\vspace{-0.4cm}
\caption{Scalability on synthetic random graphs}
\label{fig:scal1}
%\vspace{-0.4cm}
\end{figure}

\begin{figure}[h]
\begin{center}
\begin{tabular}{c}
\subfigure[\kw{Power}-\kw{law} \kw{graphs}  ($\tau$=2)]{
\includegraphics[width=0.85\columnwidth]{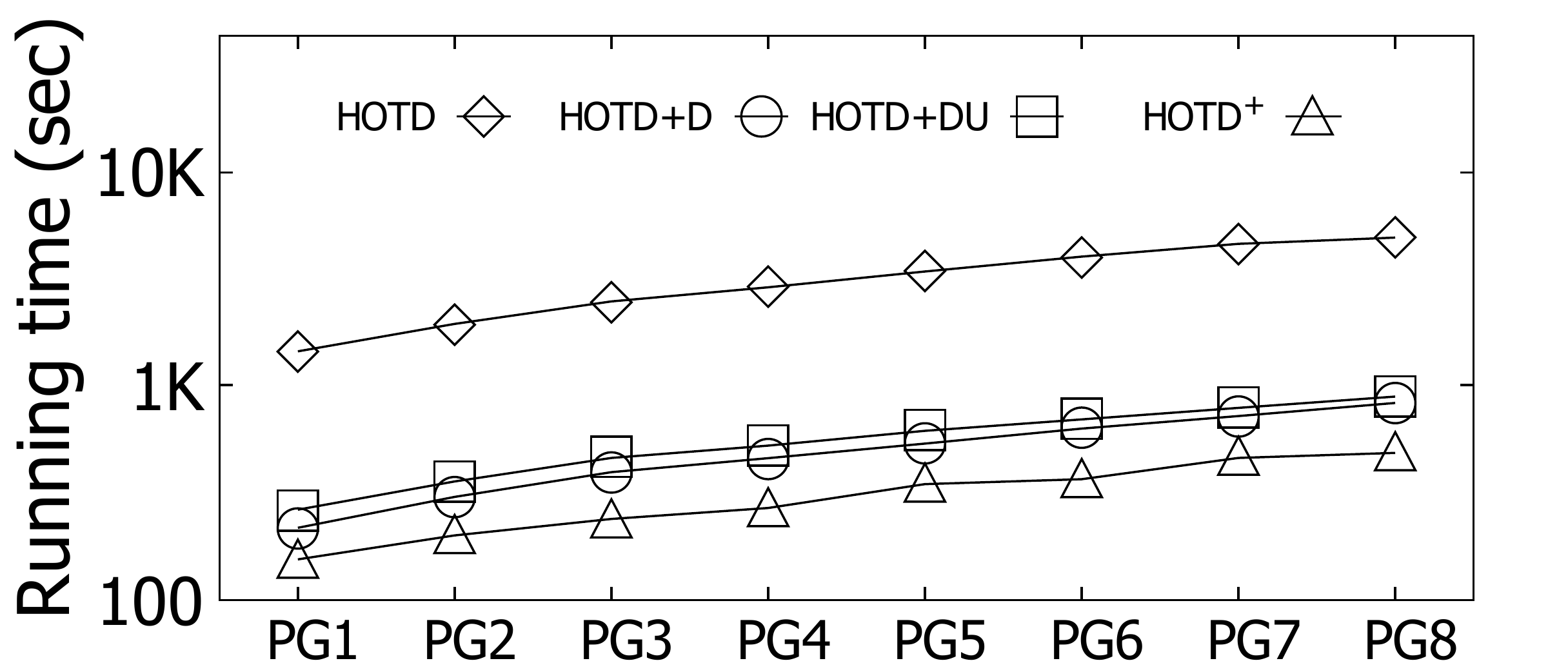}
\label{fig:scal_pg2}}
\vspace{-0.2cm}
\\
\subfigure[\kw{Power}-\kw{law} \kw{graphs}  ($\tau$=3)]{
%\hspace{-0.5cm}
\includegraphics[width=0.85\columnwidth]{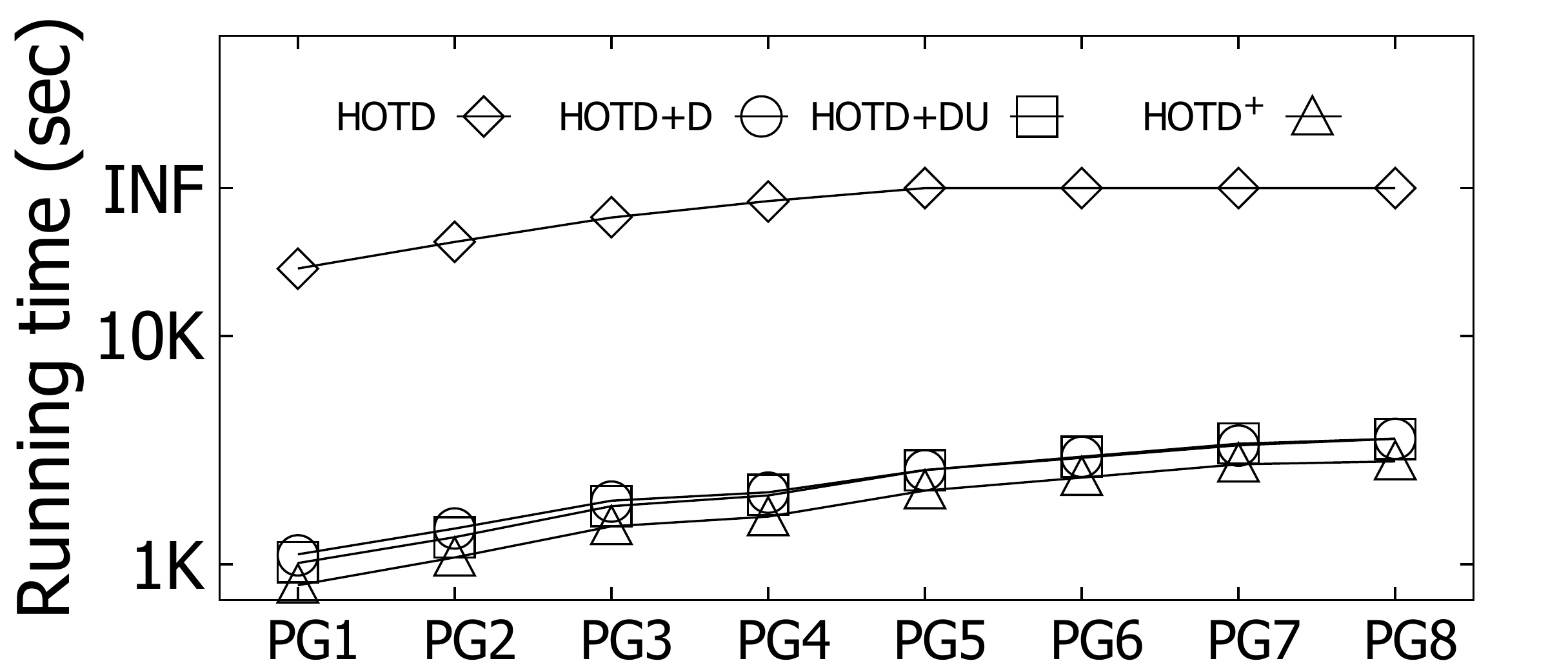}
\label{fig:scal_pg3}}
\vspace{-0.2cm}
\\
\subfigure[\kw{Power}-\kw{law} \kw{graphs}  ($\tau$=4)]{
\includegraphics[width=0.85\columnwidth]{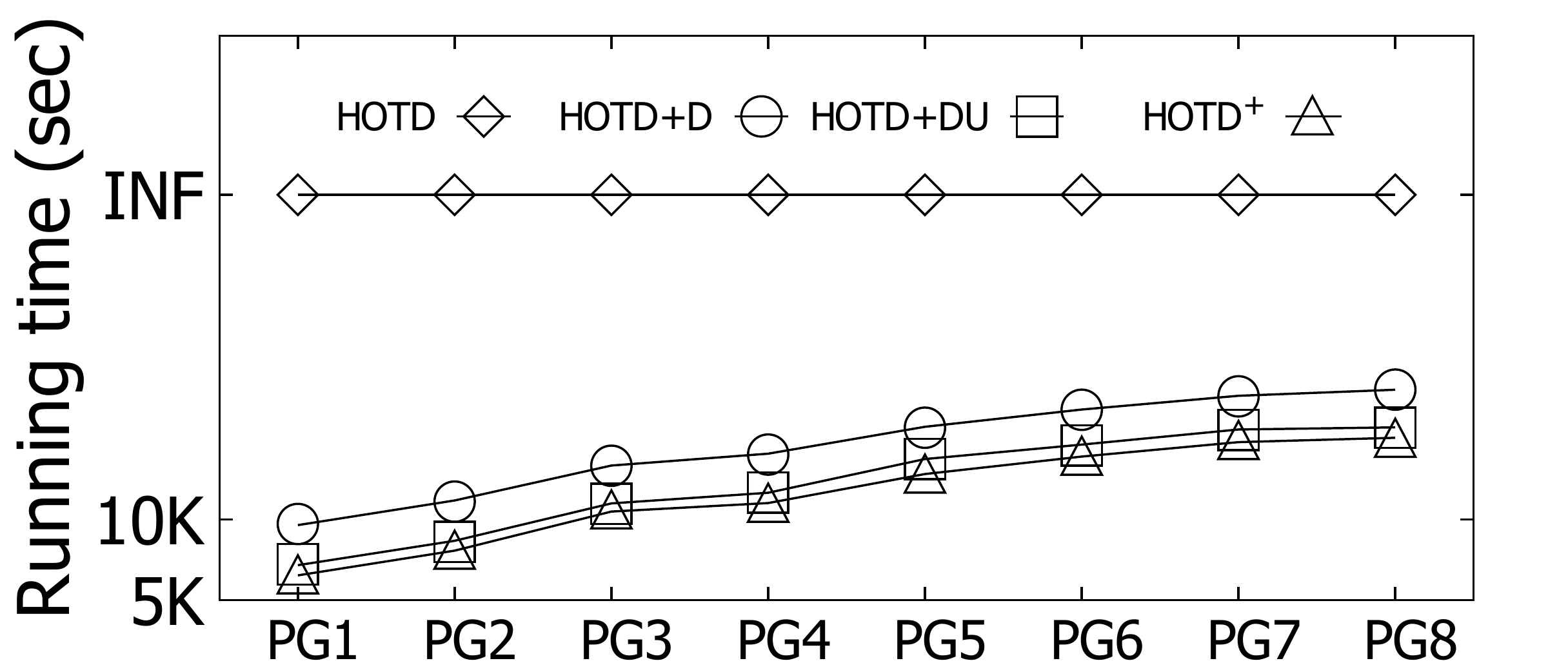}
\label{fig:scal_pg4}}

\end{tabular}
\end{center}
\vspace{-0.4cm}
\caption{Scalability on synthetic power-law graphs}
\label{fig:scal2}
\vspace{-0.4cm}
\end{figure}

%\vspace{-0.5cm}

\stitle{Exp-3: Tightness of  $\underline{\phi}_\tau(e, G)$. } In this experiment, we evaluate the tightness of the lower bound $\underline{\phi}_{\tau}(e, G)$ in \hotdp.  We use the widely adopted approximation error (AE) to measure the tightness \cite{matrixcomputation}. The approximation error for $\underline{\phi}_\tau(e, G)$ is defined as follows:

\begin{equation}
\begin{aligned}
AE(\underline{\phi}_\tau(e, G))=\frac{\sum_{i=1}^{m}{\frac{|\phi_\tau(e_i, G)-\underline{\phi}_\tau(e_i, G)|}{\phi(e_i, G)}}}{m};
\end{aligned}
\label{dif}
\end{equation}

As shown in  \reftable{lbacc}, for $\underline{\phi}_\tau(e, G)$,  when $\tau=2$,  on most datasets, the approximation error is even less than 0.1, like \kw{PT}, \kw{CG}, \kw{FB}, $\dots$. When $\tau =3$, although its performance is not as good as that of $\tau = 2$, the approximation error is still small on most datasets. This is because the $3$-hop  neighborhood information is much more complex than the $2$-hop neighborhood information. When $\tau=4$, it shows a trend similar to that of $\tau =3$. Therefore, $\underline{\phi}_\tau(e, G)$ is effective regarding $\phi_\tau(e, G)$. %Combining with the above analysis about the support computation amount, the lower bound performs well on both effectiveness and accuracy when $\tau$=2.
% However, since when $\tau>$3 there are more complex and larger neighborhood relationships between edges, the lower bound performs better when $\tau$=2.

\stitle{Exp-4: Scalability.} Finally, we evaluate the scalability of the proposed algorithms on two kinds of synthetic datasets (random graph and power-law graph). For the random graph, we generates 8 graphs varying the number of vertices from $2^{17}$ to $2^{24}$ using the method  in  \cite{DBLP:conf/ipps/HoltgreweSS10,DBLP:journals/tpds/MeyerhenkeSS17}.  For the power-law graph, we generates 8 graphs varying the number of vertices from $3M$ to $10M$  using the method in \cite{lfr} (average degree with 10, other parameters with default  setting). \reffig{scal1} and \reffig{scal2} show the running time of algorithms on random synthetic graphs and power-law synthetic graphs, respectively. 
%As shown in \reffig{scal}, the running time of the algorithms increases when the number of vertices increases. \hotdp is the most efficient one among all algorithms and shows good scalability.

\reffig{scal1} and \reffig{scal2} show that as the graph size increases, the  running time of all algorithms increases as well. This is because as the size of graphs increases, more edges have to be processed during the decomposition. Besides, for $\tau=2,3,4$, benefited from the proposed optimization strategies, \hotdp is the most efficient one among the four algorithms and shows good scalability compared with other three algorithms.

\begin{figure}[htp]
\begin{center}
\begin{tabular}{c}
\subfigure[\kw{Highschool}]{
%\hspace{-0.5cm}
\includegraphics[width=0.7\columnwidth]{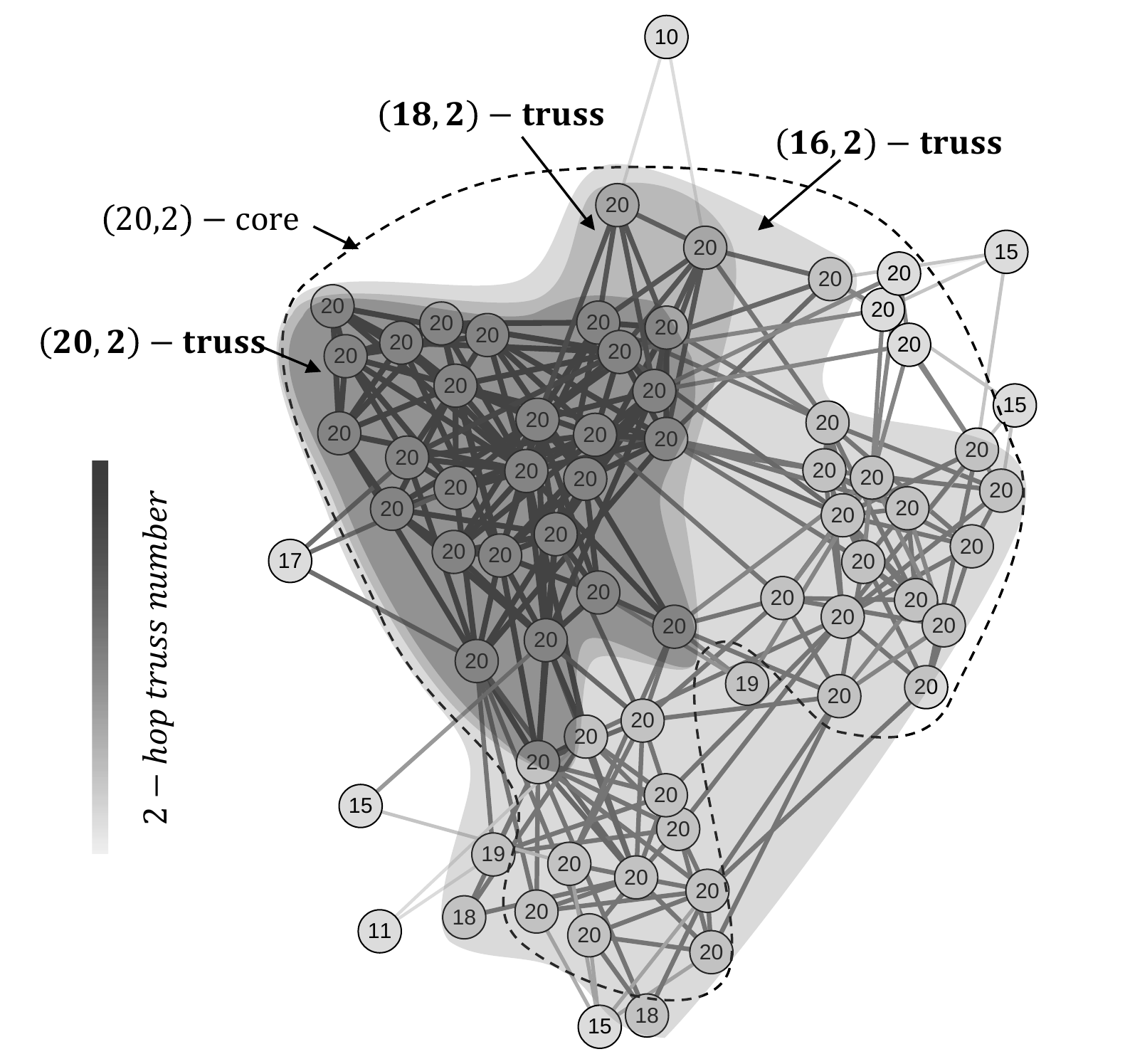}
\label{fig:case1}}
\vspace{-0.2cm}
\\
\subfigure[\kw{Political} \kw{Books}]{
\includegraphics[width=0.7\columnwidth]{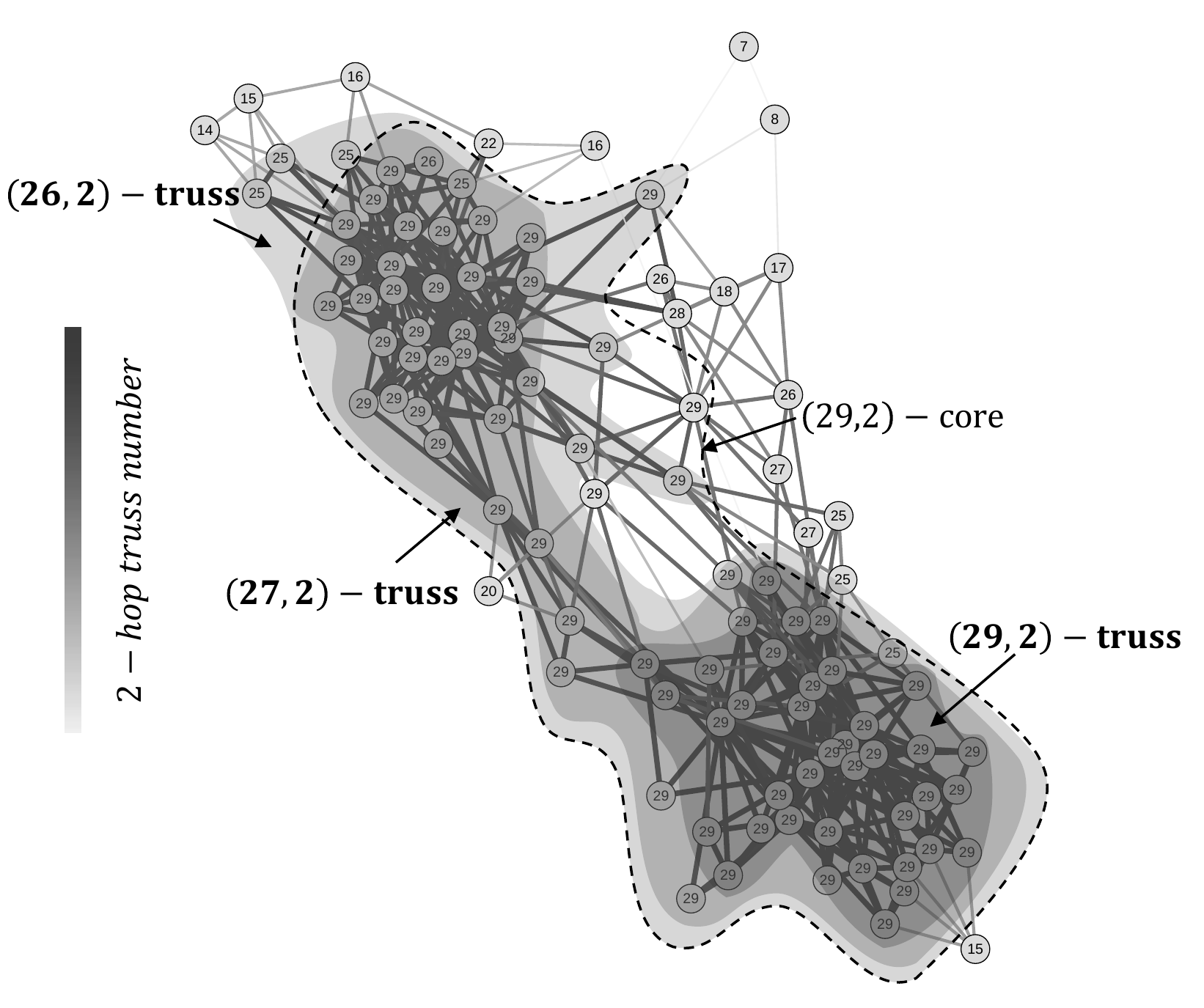}
\label{fig:case2}}

\end{tabular}
\end{center}
\vspace{-0.4cm}
\caption{Compare ($k$,$\tau$)-truss with ($k$,$h$)-core, hop=2}
\label{fig:case}
\vspace{-0.4cm}
\end{figure}

\stitle{Exp-5: Case study.} In this experiment, we compare ($k$,$\tau$)-truss with ($k$,$h$)-core which is the existing most similar cohesive subgraph model with our model. \reffig{case} shows the differences between them on two real-world graphs \kw{Highschool} and \kw{Polotical} \kw{Books} from KONECT. \kw{Highschool} is a network contains friendships between boys in a small highschool. \kw{Polotical} \kw{Books} is a network of books, edges between books represent frequent co-purchasing of books by the same buyers. As shown at \reffig{case1}, for ($k$,$h$)-core, most vertices are covered by (20,2)-core, it is hard to further distinguish more cohesive structure from it. For our model, (20,2)-core is further decomposed into more fine-grained hierarchy structure, like (16,2)-truss, (18,2)-truss and (20,2)-truss. Obviously, (20,2)-truss is the most cohesive subgraph through graph visualization. \reffig{case2} shows the similar phenomenon on \kw{Polotical} \kw{Books} dataset.  It's because a ($k$+1,$\tau$)-truss is a ($k$,$h$)-core but not vice versa, $\tau$=$h$. Therefore,  our model can further search ``core" of a ($k$,$h$)-core benefited from the more rigorous requirement for cohesiveness. In a result, our ($k$,$\tau$)-truss has higher ability to reveal fine-grained structure information.

\section{Related Work}
\label{sec:related}
\stitle{Truss decomposition.} In the literature, plenty of research efforts have been devoted to the cohesive subgraph models for graph structure analysis\cite{luce1950connectivity,seidman1983network,DBLP:conf/sigmod/BonchiKS19,truss,abello2002massive,pei2005mining,DBLP:journals/pvldb/WangZTT10,DBLP:conf/edbt/ZhouLYLCL12,DBLP:conf/icde/WenQ0CC19}. Among these cohesive subgraph models, the $k$-truss model has received considerable attention. $k$-truss  model is first introduced in \cite{truss}. In \cite{truss}, an in-memory algorithm to detect the $k$-truss for a given $k$ with time complexity $O(\sum_{v \in V(G)}d^2_1(v, G))$ is proposed. \cite{DBLP:conf/sigmod/ShaoCC14} studies the problem of detecting the $k$-truss for a given $k$ in distribute environments.  By using the triangle counting techniques proposed in \cite{DBLP:journals/tcs/Latapy08},  \cite{DBLP:journals/pvldb/WangC12}  proposes an truss decomposition algorithm with time complexity $O(m^\frac{3}{2})$.  \cite{DBLP:journals/pvldb/WangC12} also investigates  the external-memory algorithms to conduct the truss decomposition. \cite{DBLP:conf/sigmod/0001Y19} proposes an efficient algorithm to maintain the truss decomposition results in evolving graphs. \cite{DBLP:conf/sigmod/HuangLL16} investigates the truss decomposition problem in probabilistic graphs.  Compared with our $(k, \tau)$-truss model, all these models only consider the direct neighbors of an edge and the higher-order neighborhood information is missed.

\stitle{Distance-generalized cohesive subgraph models.} Clique \cite{luce1950connectivity}, $k$-core \cite{seidman1983network,DBLP:conf/sigmod/BonchiKS19} and $k$-truss \cite{truss} are three most fundamental cohesive subgraph models. For clique model, \cite{h-clique,DBLP:journals/ol/MoradiB18} propose two distance-generalized clique models, named $h$-club and $h$-clique. An $h$-club is a maximal subgraph $C$, s.t., the length of the shortest path between any two vertices of $C$ is not greater than $h$. The only difference between $h$-club and $h$-clique is that the shcortest path is within an $h$-club while it is not strictly necessary for $h$-clique. For $k$-core model, \cite{khcore} proposes ($k$,$h$)-core model which requires each vertex in it has more than $k$ neighbors within $h$-hop neighborhood. \cite{DBLP:conf/sigmod/BonchiKS19} proposes the ($k$,$h$)-core decomposition algorithm with better performance compared with \cite{khcore}. However, as shown at the experiments part (Exp-5), our model can search "core" of ($k$,$h$)-core with higher ability to reveal fine-grained structure information. Besides, there is no distance-generalized model based on $k$-truss before our model.

\section{Conclusion}
\label{sec:con}

As a representative cohesive subgraph model, $k$-truss model has received considerable attention. However,  the traditional $k$-truss model ignores the higher-order neighborhood information of an edge, which limits its ability to  reveal fine-grained structure information of the graph.  Motivated by this,  in this paper, we propose the  $(k, \tau)$-truss model and study the higher-order truss decomposition problem. We first propose a bottom-up decomposition paradigm for this problem. Based on the paradigm, we further explore three optimization strategies, namely delayed update strategy, early pruning strategy, and unchanged support detection strategy,  to improve the decomposition performance. Moreover, we also devise an efficient algorithm to compute the top $r$ ($k$,$\tau$)-trusses, which is useful in practical applications. Our experimental results on real datasets and synthetic datasets show the efficiency, effectiveness and scalability of the proposed algorithms.

\bibliographystyle{named}
\bibliography{reference}

\begin{thebibliography}{}

\bibitem[\protect\citeauthoryear{Abello \bgroup \em et al.\egroup
  }{2002}]{abello2002massive}
James Abello, Mauricio~GC Resende, and Sandra Sudarsky.
\newblock Massive quasi-clique detection.
\newblock In {\em LATIN 2002: Theoretical Informatics}, pages 598--612. 2002.

\bibitem[\protect\citeauthoryear{Abu{-}El{-}Haija \bgroup \em et al.\egroup
  }{2019}]{DBLP:conf/icml/Abu-El-HaijaPKA19}
Sami Abu{-}El{-}Haija, Bryan Perozzi, Amol Kapoor, Nazanin Alipourfard,
  Kristina Lerman, Hrayr Harutyunyan, Greg~Ver Steeg, and Aram Galstyan.
\newblock Mixhop: Higher-order graph convolutional architectures via sparsified
  neighborhood mixing.
\newblock In {\em Proceedings of {ICML}}, pages 21--29, 2019.

\bibitem[\protect\citeauthoryear{Akbas and
  Zhao}{2017}]{DBLP:journals/pvldb/AkbasZ17}
Esra Akbas and Peixiang Zhao.
\newblock Truss-based community search: a truss-equivalence based indexing
  approach.
\newblock {\em Proceedings of {VLDB} Endow.}, 10(11):1298--1309, 2017.

\bibitem[\protect\citeauthoryear{Andrade \bgroup \em et al.\egroup
  }{2006}]{PhysRevE.73.046101}
Roberto F.~S. Andrade, Jos\'e G.~V. Miranda, and Thierry~Petit Lob\~ao.
\newblock Neighborhood properties of complex networks.
\newblock {\em Phys. Rev. E}, 73:046101, Apr 2006.

\bibitem[\protect\citeauthoryear{Andrade \bgroup \em et al.\egroup
  }{2008}]{andrade2008characterization}
Roberto~FS Andrade, Jos{\'e}~GV Miranda, Suani~TR Pinho, and Thierry~Petit
  Lobao.
\newblock Characterization of complex networks by higher order neighborhood
  properties.
\newblock {\em The European Physical Journal B}, 61(2):247--256, 2008.

\bibitem[\protect\citeauthoryear{Batagelj and Zaveršnik}{2011}]{khcore}
V.~Batagelj and M~Zaveršnik.
\newblock Fast algorithms for determining (generalized) core groups in social
  networks.
\newblock {\em Adv Data Anal Classif}, 5:129--145, 2011.

\bibitem[\protect\citeauthoryear{Bonchi \bgroup \em et al.\egroup
  }{2019}]{DBLP:conf/sigmod/BonchiKS19}
Francesco Bonchi, Arijit Khan, and Lorenzo Severini.
\newblock Distance-generalized core decomposition.
\newblock In {\em Proceedings of {SIGMOD}}, pages 1006--1023, 2019.

\bibitem[\protect\citeauthoryear{Chang and Qin}{2018}]{cohesivesubgraphbook}
Lijun Chang and Lu~Qin.
\newblock {\em Cohesive Subgraph Computation over Large Sparse Graphs
  Algorithms, Data Structures, and Programming Techniques}.
\newblock Springer, 2018.

\bibitem[\protect\citeauthoryear{Chang \bgroup \em et al.\egroup
  }{2013}]{DBLP:conf/sigmod/ChangYQLLL13}
Lijun Chang, Jeffrey~Xu Yu, Lu~Qin, Xuemin Lin, Chengfei Liu, and Weifa Liang.
\newblock Efficiently computing k-edge connected components via graph
  decomposition.
\newblock In {\em Proceedings of the {SIGMOD}}, pages 205--216, 2013.

\bibitem[\protect\citeauthoryear{Chen \bgroup \em et al.\egroup
  }{2020}]{DBLP:conf/www/ChenY0QY20}
Zi~Chen, Long Yuan, Xuemin Lin, Lu~Qin, and Jianye Yang.
\newblock Efficient maximal balanced clique enumeration in signed networks.
\newblock In {\em {WWW}}, pages 339--349. {ACM} / {IW3C2}, 2020.

\bibitem[\protect\citeauthoryear{Cohen}{2008}]{truss}
J.~Cohen.
\newblock Trusses: Cohesive subgraphs for social network analysis.
\newblock In {\em {Technical report, National Security Agency}}, 2008.

\bibitem[\protect\citeauthoryear{Colomer-de Sim{\'o}n \bgroup \em et al.\egroup
  }{2013}]{colomer2013deciphering}
Pol Colomer-de Sim{\'o}n, M~Angeles Serrano, Mariano~G Beir{\'o}, J~Ignacio
  Alvarez-Hamelin, and Mari{\'a}n Bogun{\'a}.
\newblock Deciphering the global organization of clustering in real complex
  networks.
\newblock {\em Scientific reports}, 3:2517, 2013.

\bibitem[\protect\citeauthoryear{{Eades} \bgroup \em et al.\egroup
  }{2017}]{JGAA-405}
{Peter} {Eades}, {Seok-Hee} {Hong}, {An} {Nguyen}, and {Karsten} {Klein}.
\newblock Shape-based quality metrics for large graph visualization.
\newblock {\em Journal of Graph Algorithms and Applications}, 21(1):29--53,
  2017.

\bibitem[\protect\citeauthoryear{Ellson \bgroup \em et al.\egroup
  }{2002}]{10.1007/3-540-45848-4_57}
John Ellson, Emden Gansner, Lefteris Koutsofios, Stephen~C. North, and Gordon
  Woodhull.
\newblock Graphviz--- open source graph drawing tools.
\newblock In Petra Mutzel, Michael J{\"u}nger, and Sebastian Leipert, editors,
  {\em Graph Drawing}, pages 483--484, Berlin, Heidelberg, 2002. Springer
  Berlin Heidelberg.

\bibitem[\protect\citeauthoryear{Golub and Loan}{1996}]{matrixcomputation}
Gene~H. Golub and Charles F.~Van Loan.
\newblock {\em Matrix Computations}.
\newblock The Johns Hopkins University Press, 1996.

\bibitem[\protect\citeauthoryear{Holtgrewe \bgroup \em et al.\egroup
  }{2010}]{DBLP:conf/ipps/HoltgreweSS10}
Manuel Holtgrewe, Peter Sanders, and Christian Schulz.
\newblock Engineering a scalable high quality graph partitioner.
\newblock In {\em 24th {IEEE} International Symposium on Parallel and
  Distributed Processing, {IPDPS} 2010, Atlanta, Georgia, USA, 19-23 April 2010
  - Conference Proceedings}, pages 1--12. {IEEE}, 2010.

\bibitem[\protect\citeauthoryear{Huang \bgroup \em et al.\egroup
  }{2014}]{DBLP:conf/sigmod/HuangCQTY14}
Xin Huang, Hong Cheng, Lu~Qin, Wentao Tian, and Jeffrey~Xu Yu.
\newblock Querying k-truss community in large and dynamic graphs.
\newblock In {\em Proceedings of {SIGMOD}}, pages 1311--1322, 2014.

\bibitem[\protect\citeauthoryear{Huang \bgroup \em et al.\egroup
  }{2016}]{DBLP:conf/sigmod/HuangLL16}
Xin Huang, Wei Lu, and Laks V.~S. Lakshmanan.
\newblock Truss decomposition of probabilistic graphs: Semantics and
  algorithms.
\newblock In {\em Proceedings of {SIGMOD}}, pages 77--90, 2016.

\bibitem[\protect\citeauthoryear{Lancichinetti \bgroup \em et al.\egroup
  }{2008}]{lfr}
Andrea Lancichinetti, Santo Fortunato, and Filippo Radicchi.
\newblock Benchmark graphs for testing community detection algorithms.
\newblock In {\em Phys. Rev. E 78}, 2008.

\bibitem[\protect\citeauthoryear{Latapy}{2008}]{DBLP:journals/tcs/Latapy08}
Matthieu Latapy.
\newblock Main-memory triangle computations for very large (sparse (power-law))
  graphs.
\newblock {\em Theory Computer Science}, 407(1-3):458--473, 2008.

\bibitem[\protect\citeauthoryear{Lee \bgroup \em et al.\egroup
  }{2010}]{DBLP:series/ads/LeeRJA10}
Victor~E. Lee, Ning Ruan, Ruoming Jin, and Charu~C. Aggarwal.
\newblock A survey of algorithms for dense subgraph discovery.
\newblock In {\em Managing and Mining Graph Data}, pages 303--336. 2010.

\bibitem[\protect\citeauthoryear{Liu \bgroup \em et al.\egroup
  }{2019a}]{DBLP:conf/www/LiuYLQZZ19}
Boge Liu, Long Yuan, Xuemin Lin, Lu~Qin, Wenjie Zhang, and Jingren Zhou.
\newblock Efficient ({\(\alpha\)}, {\(\beta\)})-core computation: an
  index-based approach.
\newblock In {\em {WWW}}, pages 1130--1141. {ACM}, 2019.

\bibitem[\protect\citeauthoryear{Liu \bgroup \em et al.\egroup
  }{2019b}]{DBLP:journals/corr/abs-1911-04129}
Songtao Liu, Lingwei Chen, Hanze Dong, Zihao Wang, Dinghao Wu, and Zengfeng
  Huang.
\newblock Higher-order weighted graph convolutional networks.
\newblock {\em CoRR}, abs/1911.04129, 2019.

\bibitem[\protect\citeauthoryear{Liu \bgroup \em et al.\egroup
  }{2020a}]{DBLP:journals/vldb/LiuYLQZZ20}
Boge Liu, Long Yuan, Xuemin Lin, Lu~Qin, Wenjie Zhang, and Jingren Zhou.
\newblock Efficient ({\(\alpha\)}, {\(\beta\)})-core computation in bipartite
  graphs.
\newblock {\em {VLDB} J.}, 29(5):1075--1099, 2020.

\bibitem[\protect\citeauthoryear{Liu \bgroup \em et al.\egroup
  }{2020b}]{DBLP:conf/sigmod/LiuZ0XG20}
Qing Liu, Minjun Zhao, Xin Huang, Jianliang Xu, and Yunjun Gao.
\newblock Truss-based community search over large directed graphs.
\newblock In {\em Proceedings of {SIGMOD}}, pages 2183--2197, 2020.

\bibitem[\protect\citeauthoryear{Luce and Perry}{1949}]{luce1949method}
R~Duncan Luce and Albert~D Perry.
\newblock A method of matrix analysis of group structure.
\newblock {\em Psychometrika}, 14(2):95--116, 1949.

\bibitem[\protect\citeauthoryear{Luce}{1950a}]{luce1950connectivity}
R~Duncan Luce.
\newblock Connectivity and generalized cliques in sociometric group structure.
\newblock {\em Psychometrika}, 15(2):169--190, 1950.

\bibitem[\protect\citeauthoryear{Luce}{1950b}]{h-clique}
R.D Luce.
\newblock Connectivity and generalized cliques in sociometric group structure.
\newblock {\em Psychometrika}, 15:169--190, 1950.

\bibitem[\protect\citeauthoryear{Meyerhenke \bgroup \em et al.\egroup
  }{2017}]{DBLP:journals/tpds/MeyerhenkeSS17}
Henning Meyerhenke, Peter Sanders, and Christian Schulz.
\newblock Parallel graph partitioning for complex networks.
\newblock {\em {IEEE} Trans. Parallel Distributed Syst.}, 28(9):2625--2638,
  2017.

\bibitem[\protect\citeauthoryear{Mones \bgroup \em et al.\egroup
  }{2012}]{10.1371/journal.pone.0033799}
Enys Mones, Lilla Vicsek, and Tam{\'{a}}s Vicsek.
\newblock Hierarchy measure for complex networks.
\newblock {\em PLOS ONE}, 7, 03 2012.

\bibitem[\protect\citeauthoryear{Moradi and
  Balasundaram}{2018}]{DBLP:journals/ol/MoradiB18}
Esmaeel Moradi and Balabhaskar Balasundaram.
\newblock Finding a maximum k-club using the k-clique formulation and canonical
  hypercube cuts.
\newblock {\em Optim. Lett.}, 12(8):1947--1957, 2018.

\bibitem[\protect\citeauthoryear{Orsini \bgroup \em et al.\egroup
  }{2013}]{orsini2013evolution}
Chiara Orsini, Enrico Gregori, Luciano Lenzini, and Dmitri Krioukov.
\newblock Evolution of the internet $ k $-dense structure.
\newblock {\em IEEE/ACM Transactions on Networking}, 22(6):1769--1780, 2013.

\bibitem[\protect\citeauthoryear{Ouyang \bgroup \em et al.\egroup
  }{2020}]{DBLP:journals/pvldb/OuyangYQCZL20}
Dian Ouyang, Long Yuan, Lu~Qin, Lijun Chang, Ying Zhang, and Xuemin Lin.
\newblock Efficient shortest path index maintenance on dynamic road networks
  with theoretical guarantees.
\newblock {\em Proc. {VLDB} Endow.}, 13(5):602--615, 2020.

\bibitem[\protect\citeauthoryear{Pei \bgroup \em et al.\egroup
  }{2005}]{pei2005mining}
Jian Pei, Daxin Jiang, and Aidong Zhang.
\newblock On mining cross-graph quasi-cliques.
\newblock In {\em Proceedings of SIGKDD}, pages 228--238, 2005.

\bibitem[\protect\citeauthoryear{Sahu \bgroup \em et al.\egroup
  }{2017}]{DBLP:journals/pvldb/SahuMSLO17}
Siddhartha Sahu, Amine Mhedhbi, Semih Salihoglu, Jimmy Lin, and M.~Tamer
  {\"{O}}zsu.
\newblock The ubiquity of large graphs and surprising challenges of graph
  processing.
\newblock {\em Proc. {VLDB} Endow.}, 11(4):420--431, 2017.

\bibitem[\protect\citeauthoryear{Seidman}{1983}]{seidman1983network}
Stephen~B Seidman.
\newblock Network structure and minimum degree.
\newblock {\em Social Networks}, 5(3):269--287, 1983.

\bibitem[\protect\citeauthoryear{Shao \bgroup \em et al.\egroup
  }{2014}]{DBLP:conf/sigmod/ShaoCC14}
Yingxia Shao, Lei Chen, and Bin Cui.
\newblock Efficient cohesive subgraphs detection in parallel.
\newblock In {\em Proceedings of {SIGMOD}}, pages 613--624, 2014.

\bibitem[\protect\citeauthoryear{Sun \bgroup \em et al.\egroup
  }{2020}]{DBLP:conf/aaai/SunW0CDZQ20}
Zequn Sun, Chengming Wang, Wei Hu, Muhao Chen, Jian Dai, Wei Zhang, and Yuzhong
  Qu.
\newblock Knowledge graph alignment network with gated multi-hop neighborhood
  aggregation.
\newblock In {\em Proceedings of {AAAI}}, pages 222--229, 2020.

\bibitem[\protect\citeauthoryear{Wang and
  Cheng}{2012}]{DBLP:journals/pvldb/WangC12}
Jia Wang and James Cheng.
\newblock Truss decomposition in massive networks.
\newblock {\em Proceedings of {VLDB} Endow.}, 5(9):812--823, 2012.

\bibitem[\protect\citeauthoryear{Wang \bgroup \em et al.\egroup
  }{2010}]{DBLP:journals/pvldb/WangZTT10}
Nan Wang, Jingbo Zhang, Kian-Lee Tan, and Anthony K.~H. Tung.
\newblock On triangulation-based dense neighborhood graphs discovery.
\newblock {\em Proceedings of {VLDB} Endow.}, 4(2):58--68, 2010.

\bibitem[\protect\citeauthoryear{Wen \bgroup \em et al.\egroup
  }{2019}]{DBLP:conf/icde/WenQ0CC19}
Dong Wen, Lu~Qin, Ying Zhang, Lijun Chang, and Ling Chen.
\newblock Enumerating k-vertex connected components in large graphs.
\newblock In {\em Proceedings of {ICDE}}, pages 52--63, 2019.

\bibitem[\protect\citeauthoryear{Wu \bgroup \em et al.\egroup
  }{}]{DBLP:conf/dasfaa/WuYLYZ19}
Xudong Wu, Long Yuan, Xuemin Lin, Shiyu Yang, and Wenjie Zhang.
\newblock Towards efficient k-tripeak decomposition on large graphs.
\newblock In {\em {DASFAA}}.

\bibitem[\protect\citeauthoryear{Xue \bgroup \em et al.\egroup
  }{2020}]{DBLP:conf/bigcomp/XueSS20}
Hui Xue, Xin{-}Kai Sun, and Wei{-}Xiang Sun.
\newblock Multi-hop hierarchical graph neural networks.
\newblock In {\em Proceedings of BigComp}, pages 82--89, 2020.

\bibitem[\protect\citeauthoryear{Yuan \bgroup \em et al.\egroup
  }{2016a}]{DBLP:journals/vldb/YuanQLCZ16}
Long Yuan, Lu~Qin, Xuemin Lin, Lijun Chang, and Wenjie Zhang.
\newblock Diversified top-k clique search.
\newblock {\em {VLDB} J.}, 25(2):171--196, 2016.

\bibitem[\protect\citeauthoryear{Yuan \bgroup \em et al.\egroup
  }{2016b}]{DBLP:journals/pvldb/YuanQLCZ16}
Long Yuan, Lu~Qin, Xuemin Lin, Lijun Chang, and Wenjie Zhang.
\newblock {I/O} efficient {ECC} graph decomposition via graph reduction.
\newblock {\em Proc. {VLDB} Endow.}, 9(7):516--527, 2016.

\bibitem[\protect\citeauthoryear{Yuan \bgroup \em et al.\egroup
  }{2017}]{DBLP:journals/pvldb/YuanQLCZ17}
Long Yuan, Lu~Qin, Xuemin Lin, Lijun Chang, and Wenjie Zhang.
\newblock Effective and efficient dynamic graph coloring.
\newblock {\em Proc. {VLDB} Endow.}, 11(3):338--351, 2017.

\bibitem[\protect\citeauthoryear{Yuan \bgroup \em et al.\egroup
  }{2018}]{DBLP:journals/tkde/YuanQZCY18}
Long Yuan, Lu~Qin, Wenjie Zhang, Lijun Chang, and Jianye Yang.
\newblock Index-based densest clique percolation community search in networks.
\newblock {\em {IEEE} Trans. Knowl. Data Eng.}, 30(5):922--935, 2018.

\bibitem[\protect\citeauthoryear{Zhang and Yu}{2019}]{DBLP:conf/sigmod/0001Y19}
Yikai Zhang and Jeffrey~Xu Yu.
\newblock Unboundedness and efficiency of truss maintenance in evolving graphs.
\newblock In {\em Proceedings of {SIGMOD}}, pages 1024--1041, 2019.

\bibitem[\protect\citeauthoryear{Zhou \bgroup \em et al.\egroup
  }{2012}]{DBLP:conf/edbt/ZhouLYLCL12}
Rui Zhou, Chengfei Liu, Jeffrey~Xu Yu, Weifa Liang, Baichen Chen, and Jianxin
  Li.
\newblock Finding maximal k-edge-connected subgraphs from a large graph.
\newblock In {\em Proceedings of {EDBT}}, pages 480--491, 2012.

\end{thebibliography}
%\newpage
%\input{sup}
\end{document}